\documentclass[10pt]{article}

\usepackage{amsmath}
\usepackage{amssymb}

\usepackage{graphicx}

\usepackage{cite}

\usepackage{color} 

\usepackage[labelfont=bf,labelsep=period,justification=raggedright]{caption}

\makeatletter
\renewcommand{\@biblabel}[1]{\quad#1.}
\makeatother

\date{}

\newcommand{\spstart}{\text{start}}
\newcommand{\spend}{\text{end}}
\newcommand{\D}{\mathcal{D}}

\begin{document}










\title{Computational Inference Methods for HIV-1 Selective Sweeps Shaped by Early Cytotoxic T-Lymphocyte Response}
\author{Sivan Leviyang\thanks{
Department of Mathematics and Statistics, Georgetown University, Washington, DC, USA.
E-mail: sr286@georgetown.edu}}
\maketitle

\abstract{
Cytotoxic T lymphocytes (CTLs) play an important role in shaping HIV-1 infection.   In particular, during the first weeks of infection, CTLs select for multiple escape mutations in the infecting HIV population.  In recent years, methods have been developed that use intra-patient escape mutation data to estimate rates of escape from CTL selection. The resultant escape rate estimates have been used to study CTL kill rates and fitness costs associated with particular mutations, thereby providing a quantitative framework for exploring CTL response and HIV dynamics using patient datasets.    Current methods for escape rate inference focus on a specific HIV mutant selected by a single CTL response.  However, recent studies have shown that during the first weeks of infection, CTL responses occur at $1-3$ epitopes and HIV escape occurs through complex mutation pathways.   In this work we develop a model of initial infection, based on the well-known standard model, that allows us to model the complex mutation pathways of HIV escape.  Under this model, we develop two computational inference methods.  In one, we use a Bayesian approach to construct posteriors for the parameters of our model.  In the second, we develop methods for hypothesis testing under our model.   The methods are applied to two CHAVI datasets, demonstrating the importance of taking into account the interaction of multiple mutant variants and multi-directional selection.  }

\section{Introduction}

	Acute HIV-1 infection is marked by a period of approximately $2-4$ weeks in which the viral population expands from a presumed $1-5$ initial infected cells, in the case of sexual transmission \cite{Keele_2008_PNAS}, to a population of approximately $10^9$ infected cells \cite{Fiebig_2003_AIDS, Mehandru_2004_JEM, Mehandru_2007_J_Virol, Cohen_2011_NJM}.   An innate immune response occurs within about $5$ days of initial infection, but the adaptive response takes more time.   In particular, cellular response is correlated temporally with the end of the expansion period. \cite{Cohen_2011_NJM, Borrow_1994_J_Virol, Koop_1994_J_Virol}. 
	
	Adaptive response during the acute phase of infection includes both a T cell and antibody component, but neutralizing antibodies only arise towards the end of the acute phase or later in infection \cite{Tomaras_2008_J_Virol, Bar_2012_PLOS_Pathogens}.  In contrast, there is strong evidence suggesting that CTLs do exert a significant selective force around the time of peak viral load \cite{Fernandex_J_Virol_2005, Goonetilleke_2009_JEM}.  For example, in the four patients considered in \cite{Goonetilleke_2009_JEM}, initial CTL response targeting a single epitope started at or several days before peak viral load and escape mutations at the epitope were fixed or nearly so $2-3$ weeks thereafter.  Shortly after the initial CTL response, CTL responses at $1-2$ other epitopes was seen.  Escape mutations at these epitopes were fixed or nearly so within roughly $4-6$ weeks of peak viral load.  
	
	While the description above provides a qualitative picture of initial CTL response and HIV escape, an accompanying quantitative description is still being developed.   Put in evolutionary terms, HIV escape from CTL response forms an example of a selective sweep.  A quantitative theory of selective sweeps does exist, e.g. \cite{Gillespie_1991_book, Kaplan_Genetics_1988, Krone_TPB_1997, Nielson_1998_Genetics, Tajima_1989_Genetics}, but this theory typically builds off generic models that do not fit the biology of acute HIV infection in several ways.   First, the HIV population size is not fixed during initial infection, instead HIV viral load rises by $2-4$ logs, peaks, and then drops by $2-3$ logs during the time span of the initial HIV selective sweep \cite{Fiebig_2003_AIDS, Goonetilleke_2009_JEM, Stafford_2000_J_Theor_Bio}.  Second, since CTL populations change over the time period of the sweep, the selective force exerted by CTLs is time varying.   Third, since CTLs target multiple epitopes and the fitness effects of mutations are complex, the selective force exerted on the HIV population is multi-directional.  Fourth, the high mutation rate of HIV means not only that many mutant variants will arise, but also that multiple mutation pathways leading to escape will exist.   As a result, providing a quantitative description of HIV selective sweeps during initial infection is a modeling and computational challenge.

	Several authors have examined selection in the context of HIV, e.g. \cite{Frost_2000_J_Virol, Nielson_1998_Genetics, Pennings_2006_PLOS_Genetics, Rouzine_2010_TPB, Batorsky_2011_PNAS}, but not with a focus on making inferences on selective sweeps in acute infection and, consequently, not with models that reflect the unique features of CTL response and HIV dynamics during acute infection.   Techniques have been developed to infer HIV escape rates by focusing on mutant frequency at a single epitope \cite{Fernandex_J_Virol_2005, Asquith_PLOS_Biology_2006, Goonetilleke_2009_JEM, Ganusov_2011_J_Virology, Ganusov_2006_PLOS_Comp_BIO}.  These techniques have been valuable in quantifying the role of CTL attack in shaping HIV escape rates.  However, the model used considers a single genome region and assumes only two HIV variants at that region, a wild type variant and a mutant variant.  As a result, the effect of different HIV variants exposed to multi-directional selection is difficult to assess.

	In this work, we describe a model and associated computational methods through which HIV selective sweeps driven by multi-epitope CTL response and multi-variant HIV escape can be analysed.   The model is based on the standard model of viral dynamics \cite{Perelson_Nature_Reviews_2002}.  Since HIV mutants that escape CTL attack do not exist at initial infection and often arise sequentially in time,  we extend the deterministic standard model to a stochastic birth-death process that includes mutation.  Intuitively, the birth-death process is an agent-based system that tracks the birth, death, and mutation events of individual cells infected by different HIV variants.   Such an approach to HIV dynamics has been used previously, e.g.
\cite{Ribeiro_AIDS_1999, Tuckwell_Math_Biosci_2008, Merrill_J_Comp_App_Math_2005}.  We model the possible mutation pathways by what we term an escape graph.   The escape graph, formed through information available in the datasets considered, specifies the different mutations and resultant HIV variants that are involved in the selective sweep and tracked by the birth-death process.
	
	 Given a model, specified through an escape graph and associated birth-death process, we develop two computational inference methods.   First, assuming a prior distribution on the parameters of the birth-death process, we describe a computational approach to forming a posterior parameter distribution.   Posteriors describe the most likely parameters conditioned on the data, but they do not directly describe the overall fit of the model to the data.   So second, we describe a computational approach that determines the p-value of a specific escape graph and associated birth-death process.  This allows for hypothesis testing.
	 
	 	We apply our computational methods to the datasets of patients CH40 and CH58 presented in \cite{Goonetilleke_2009_JEM, Fisher_2010_PLOS_One}, focusing our analysis on the first $2-3$ timepoints sampled in \cite{Goonetilleke_2009_JEM}.  Our methods use data that specifies frequencies of different HIV variants at different sampled time points, the type of data available in \cite{Goonetilleke_2009_JEM, Fisher_2010_PLOS_One}. For CH58, we focus our analysis on the first two regions of the founder genome at which mutations are seen to fix.  One region is the epitope targeted by the initial CTL response and the other region is possibly associated with fitness effects.  Mutations fix at these two regions simultaneously and we construct posteriors to analyze how these simultaneous sweeps affect each other.  For CH40, we consider escape from the initial CTL response.   Through deep sequencing data for CH40 presented in \cite{Fisher_2010_PLOS_One}, $7$ different mutations at the epitope targeted are seen to play a part in escape.   By exploiting our p-value computations to perform hypothesis tests, we show that variations in mutant frequency seen in the CH58 dataset are not necessarily reflective of difference in CTL kill rates or fitness difference, but rather may result from the stochasticity inherent in the time at which different mutant variants arise.  
	  
	  Calculating posteriors and p-values under our model is not computationally trivial.     The model is stochastic and the data is high-dimensional.   Standard Monte Carlo approaches in which the birth-death process is simulated without any conditioning are not effective because the data represents a single point in a high dimensional space and the birth-death process rarely hits any specific point.    Further, since the data is high-dimensional, defining what is meant by a p-value is not straightforward.  We address these issue by exploiting a stochastic-deterministic decomposition of HIV birth-death processes introduced and explored by several authors \cite{Rouzine_Micro_Review_2001, Nowak_Book, Desai_Genetics_2007, Leviyang_HIV_Sampling}.  Using this decomposition, we construct a reduction that associates the  birth-death process with a stochastically simpler Markov chain.   Through this Markov chain reduction, the posterior can be computed using a Markov Chain-Monte Carlo approach and the p-value can be defined and computed.


\section{Model and Methods}

 	Our model is based on the following form of the standard model for which virions are assumed to be in steady state,  \cite{Perelson_Nature_Reviews_2002, Nowak_and_May_Book}
\begin{align}  \label{E:standard_model}
\dot{T}(t) & = \lambda - dT - k T \sum_v I_v \\ \notag
\dot{I}_v(t) & = k T I_v - \delta_v I_v \\ \notag
\end{align}
where $T, I_v$ represent uninfected target CD4+ T cells and CD4+ T cells infected by HIV variant $v$, respectively, per $\mu$L.  To allow for multiple variants, $v$ varies over whatever set of variants is being considered.
$k, \lambda, d$ represent the infection rate per target cell per $\mu$L, target cell production rate per $\mu$L, and target cell death rate, respectively, per day.  Crucial for our model, $\delta_v$ represent the death rates of cells infected by variant $v$.   Instead of explicitly modeling CTL dynamics, we allow the parameters $\delta_v$ for different variants to vary in time and in this way model the selective force exerted by CTLs or other fitness effects implicitly.

\begin{figure} [h]
\begin{center} 
\includegraphics[width=4in]{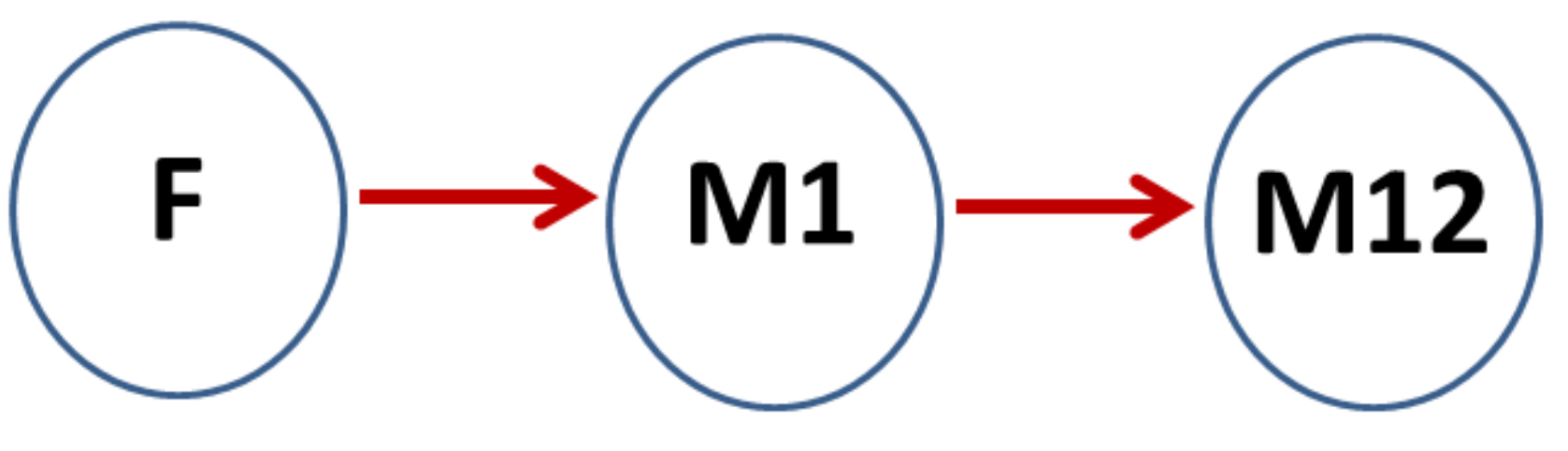}
\end{center}
\caption{\textbf{Escape Graph for CH58}} 
\label{F:CH58_escape_graph}
\end{figure}

\begin{figure} [h]
\begin{center} 
\includegraphics[width=4in]{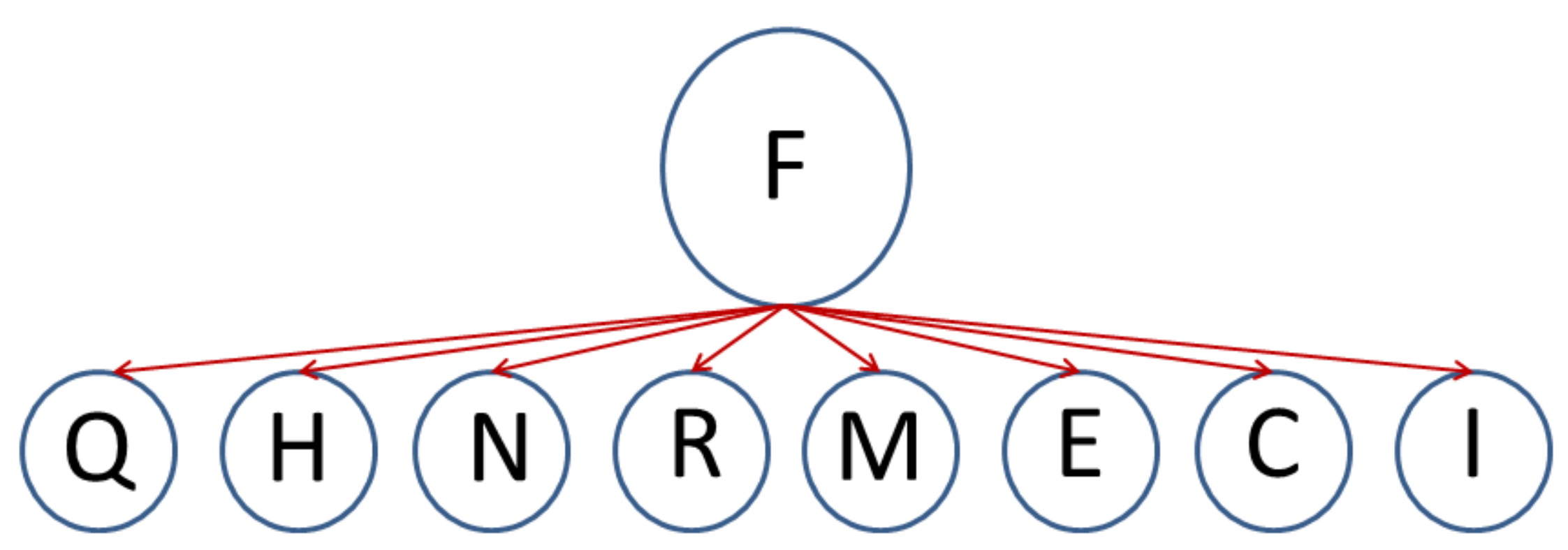}
\end{center}
\caption{\textbf{Escape Graph for CH40}} 
\label{F:CH40_escape_graph}
\end{figure}

	To model the mutation pathways through which HIV variants escape selection, we define an escape graph by specifying a set of vertices and a set of directed edges.  Vertices correspond to HIV variants that are part of the mutation pathway through which the selective sweep occurs.  Two vertices, say $A$ and $B$, may be connected by an edge directed from $A$ to $B$ if variant $A$ can mutate into variant $B$ through a single nucleotide substitution.  We write $A \to B$ when $A$ has an edge pointing to $B$.  Figure \ref{F:CH58_escape_graph} shows an escape graph with three variants : $F$, $M1$, and $M12$.  In the results section, this escape graph arises in connection to patient CH58.   $F$ represents the original founder HIV variant that infected the patient.   The movement from variant $F$ to $M1$ to $M12$ represents a pathway of HIV escape suggested by the patient data.   The specific geometry of this graph, which ignores certain variants, reflects a modeling choice.  Figure \ref{F:CH40_escape_graph} shows the escape graph we use for patient CH40.   
	
	A birth-death process describes the dynamics of the variant populations specified by the escape graph vertices and is an extension of (\ref{E:standard_model}).  $T(t)$ and $I_v(t)$ represent the same populations as in the standard model, with the caveats that the $v$'s are vertices in the escape graph and the units are now per body rather than per $\mu$L.  The change in units allows us to track the rise of new variants which initially only infect a single cell.   The parameters of the birth-death process include the parameters of (\ref{E:standard_model}) with two extensions.  First, we allow $\delta_v$ to depend on time, i.e. we consider $\delta_v(t)$ instead of just $\delta_v$.  Second, if $A$ and $B$ are connected by an edge we let $\mu_{AB}$ be the rate at which $A$ variants mutate into $B$ variants.  Throughout this work we set all such mutation rates equal to $\mu = 3 * 10^{-5}$, but the methods allow for any value for any edge.  The birth-death process is defined through the birth and death rates specified in Table \ref{T:rates}.   For example, at time $t$ a single cell infected by variant $v$ produces child infected cells with rate $kT(t)$, meaning that in a small time interval $[t, t+\Delta t]$ the probability of a new $v$ variant infected cell arising is roughly $k T(t) I_v(t) \Delta t$.   In both cases discussed in the Results section, the escape graph includes a founder vertex $F$.  We always start the birth-death process at 'initial infection' which we model as $t=0$, $I_F(0) = 1$ and $I_v(0) = 0$ for $v \ne F$.  
	
\begin{table}[h]
\begin{center}
\caption{Rates for the Birth-Death Process}
\begin{tabular}{|c||c|c|c|}  
\hline 
cell type  & birth rate & death rate & mutation rate \\
\hline
$T$  & $\lambda$ &  $d + \sum_{v} k I_v$ & -
\\
\hline
$I_v$  & $k T$ & $\delta_v(t)$ & $\mu \sum_{v' \to v} k T I_{v'}$
\\
\hline
\end{tabular}
\label{T:rates}
\end{center}
\end{table}

	Since we do not explicitly model CTL dynamics, the $\delta_v(t)$ which implicitly model CTL attack are parameters of special interest.  In computing posteriors, estimating $\delta_v(t)$ with no restrictions is beyond our current methods.  Instead, along with the escape graph and birth-death rates we specify a list of attack intervals $[0, t_1]$, $[t_1, t_2]$, \dots, $[t_{n-1}, t_n]$ and restrict the $\delta_v(t)$ to be constant during any attack interval.   For example, we often choose $t_1 = 15$ reflecting approximately $15$ days before CTL response arises and correspondingly choose $\delta_v(t) = .4$ for $t \in [0,15]$ which gives a half-life for infected cells falling between $1-2$ days \cite{Perelson_Science_1996, Stafford_2000_J_Theor_Bio}.   Posterior construction centers on computing the values of the $\delta_v(t)$ during the attack intervals.  In the Results section, $k, d, \lambda, \mu$ are fixed within each patient, although these parameters can certainly be made part of the prior and posterior.   The specific form of the attack intervals is a modeling choice and is formed by considering CTL data available.   Start and end times for the attack intervals can be made part of the posterior as is done for patient CH40.

\subsection{Simulation of the Birth-Death Process}

	Simulating a birth-death process through a Gillespie algorithm, \cite{Gillespie_2001_J_Chem_Phys}, is not computationally feasible for an infected cell population size of the order $10^9$.  Instead we employ the idea of stochastic-deterministic decomposition used by several authors \cite{Rouzine_Micro_Review_2001, Nowak_Book, Desai_Genetics_2007, Leviyang_HIV_Sampling}.  To make this decomposition precise, for each variant $v$ we define a stochastic interval, $[t_\spstart^{(v)}, t_\spend^{(v)}]$.   Given the stochastic interval, $I_v(t)$ dynamics are generated according to the following algorithm,
\begin{description}
\item[Pre-Stochastic Interval Step:] \hfill \\
For $t < t_\spstart^{(v)}$, $I_v(t) = 0$.
\item[Stochastic Interval Step:] \hfill \\
For $t \in [t_\spstart^{(v)}, t_\spend^{(v)}]$, $I_v(t)$ dynamics are generated by a Gillespie algorithm using the rates of the birth-death process.
\item[Post-Stochastic Interval Step:] \hfill \\
For $t > t_\spend^{(v)}$, $I_v(t)$ dynamics are approximated using the differential equation:
\begin{equation} \label{E:det_analogue}
\dot{I}_v = k T I_v  - \delta_v(t)I_v + \mu  \sum_{v' \to v} k T I_{v'}.
\end{equation}
\end{description}
The above algorithm can be employed for each variant, but in practice we assume that the founder variant is deterministic from $t=0$.  This approach ignores stochasticity in the first $1-2$ days of infection, a time period that does not affect our results and for which the modeling is uncertain with the advantage of improved computational time.  For all other variants, since the Gillespie algorithm samples exactly from the birth-death process, the accuracy of the stochastic-deterministic decomposition improves as the stochastic interval is widened, but at the cost of greater computation time.   The endpoints of the stochastic intervals are set through tuning parameters $\epsilon$ and $L$.   These two parameters tune the algorithm by implicitly selecting a trade-off between computational speed and accuracy. 

	To explain how we determine the stochastic interval,  we consider the escape graph in Figure \ref{F:CH58_escape_graph}.   Starting with variant $M1$, since $F$ does not have a stochastic interval,  $t_\spstart^{(M1)}$ is defined by $\mu k T(t_\spstart^{(M1)}) I_F(t_\spstart^{(M1)}) = \epsilon$, i.e. the time at which the rate of mutations reaches $\epsilon$.   The probability of an $F$ to $M1$ mutation prior to $t_\spstart^{(M1)}$ is order $\epsilon$.  The error of ignoring such mutations, as is done by the Pre-Stochastic Interval Step, drops as $\epsilon$ is decreased.  $t_\spend^{(M1)}$ is defined by  $\tilde{I}_{M1}(t_\spend^{(M1)}) = L$ where $\tilde{I}_{M1}(t)$ is roughly the average population size of $M1$ variants.  Put precisely, let $t^*$ be defined by $\mu k I_F(t^*) T(t^*) = 1$, then $\tilde{I}_{M1}(t)$ obeys (\ref{E:det_analogue}) with initial condition $\tilde{I}(t^*) = 0$.  As $L$ is increased, $t_\spend^{(M1)}$ rises thereby increasing the duration of the stochastic interval and improving accuracy.   The stochastic interval of $M12$ is generated similarly, but with $M1$ playing the role of $F$.   For a general escape graph, we would start with the founder and work our way outwards along the edges, generating a stochastic interval for each vertex.  
	
\begin{figure} [!ht]
\begin{center} 
\includegraphics[width=4in]{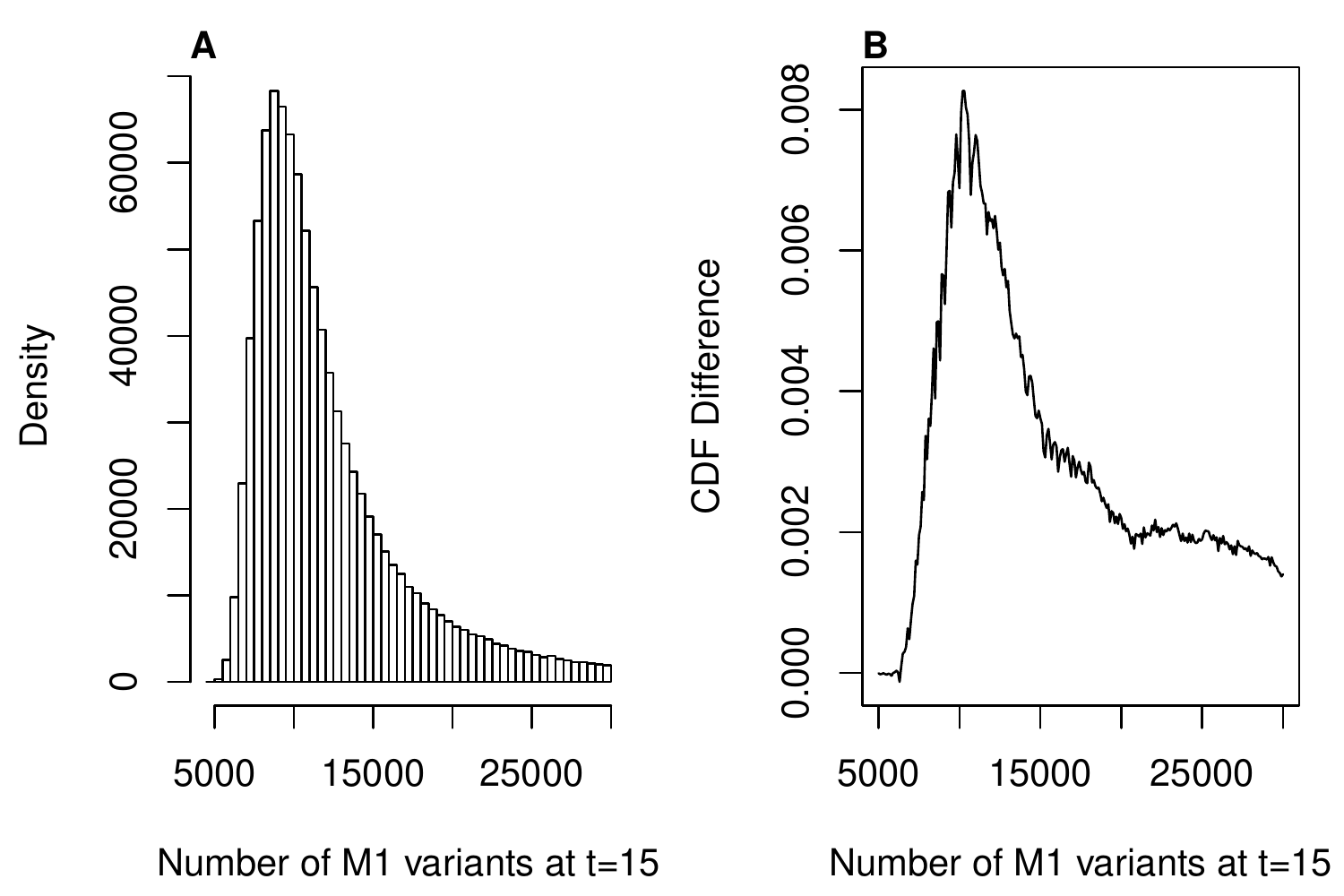}
\end{center}
\caption{\textbf{Accuracy of Stochastic-Determinsitic Decomposition}.  (A) Distribution of $I_{M1}(15)$, the number of $M1$ variants at $t=15$, with parameters $k = 2.6*10^{-3}, d = .01, \lambda = 10^8$ and $\delta_v = .4$ for $v=F,M1,M12$.     (B) Difference of cumulative distributions for $I_{M1}(15)$ using Gillespie simulation and stochastic-deterministic decomposition with $L=2000$.   Both (A) and (B) were generated using $10^6$ simulations.} 
\label{F:L_2000}
\end{figure}

	In \cite{Leviyang_HIV_Sampling}, through simulation the relative error produced by $L=100$, $\epsilon=0$ was found to be approximately $.03$.  Similar results, expressed in somewhat different settings, were found in \cite{Desai_Genetics_2007, Rouzine_Micro_Review_2001}.   Figure \ref{F:L_2000}A shows the probability density of $I_{M1}(15)$, the number of $M1$ variant infected cells at $t=15$, computed using the stochastic-deterministic decomposition with $L=2000$, $\epsilon=0$.   Parameters were set at $k = 2*10^{-3}$, $d = .01$, $\lambda = 10^8$, and $\delta_v = .4$ for $v=F,M1,M12$.  These parameter choices fall within the typical range used for HIV, e.g. \cite{Stafford_2000_J_Theor_Bio, Perelson_1993_Math_BioSci, DeBoer_2007_J_Virology, Conway_2011_PLOS_Comp_Bio}, with the caveat that $\lambda, k$ must be converted to per body units and remembering that we assume virions are in steady state.   (For example, the $k=2*10^{-3}$ in our model corresponds to $k=1.2*10^{-5}$ in the model of \cite{Conway_2011_PLOS_Comp_Bio}, approximately the value cited in that paper.)    The stochastic interval for $M1$ was $[4.6, 13.7]$, so by $t=15$ the $M1$ variants were in the deterministic portion of the decomposition.  Figure \ref{F:L_2000}B shows the difference in the cumulative distribution between using $L=2000$ and an exact Gillespie distribution run up to $t=15$.     As can be seen the maximum difference in the cdfs is less than $.01$.  In the Results section, we make the conservative choice of $L=2000$, although $L=100$ is likely sufficient.  To simulate the birth-death process, choosing $\epsilon = 0$ is computationally feasible, however the Markov chain reduction depends on $\epsilon > 0$ as will be discussed below.

\subsection{Markov Chain Reduction of the Birth-Death Process}

	While simulation of the birth-death process is possible using the algorithm of the previous section, inference is difficult.   To overcome this obstacle, we reduce the stochasticity of the birth-death process to the much simpler stochasticity of a Markov chain.   The computational approaches used to construct posteriors and p-values depend on this reduction.
	
	  As a first step in obtaining the Markov chain reduction, we modify the simulation algorithm described in the previous section.  Everything is as before, except that the \textit{Stochastic Interval Step} is now implemented through the following sub-steps which do not use a Gillespie simulation.
\begin{description}
\item[Modified Stochastic Interval Step, sub-step a] \hfill \\
Set $I_v(t) = 0$ for all $t \in [t_\spstart^{(v)}, t_\spend^{(v)})$.  During this time interval,  store $T(t)$ and $I_{v'}(t)$ for all $v'$ with an edge pointing to $v$.   
\item[Modified Stochastic Interval Step, sub-step b] \hfill \\ 
Let $B(t)$ be a single variant birth-death process, i.e. $B(t)$ is a scalar, defined by $B(t_\spstart^{(v)}) = 0$ and with birth, death, and mutation rates of $kT(t)$, $\delta_v(t)$, and $\sum_{v' \to v} \mu k T(t) I_{v'}(t)$, respectively.  $B(t)$ has the same birth, death, and mutation rates as variant $v$ infected cells.  Define $X_v = B(t_\spend^{(v)})$.    Since $T(t), I_{v'}(t)$ were stored in sub-step a, the distribution of $X_v$ can be computed through standard methods, see \cite{Conway_2011_PLOS_Comp_Bio} for an example in the context of viral dynamics.  Briefly, $X_v$ is computed by solving a backwards equation for the expected value $E[\exp[-i\omega B(t_\spend^{(v)})] \ | \ B(t) = 1]$.  Then a Fourier transform in $\omega$ is performed to obtain the distribution of $B(t_\spend^{(v)})$ or in other words $X_v$.  
\item[Modified Stochastic Interval Step, sub-step c] \hfill \\
Using the distribution computed in sub-step b, produce a sample $\hat{x}_v$ from $X_v$.   Set $I_v(t_\spend^{(v)}) = \hat{x}_v$.
\end{description}
The above algorithm samples the number of variants at the end of the stochastic interval exactly, under the assumption that $v$ variant dynamics during the stochastic interval do not effect $T(t)$ dynamics, i.e. the number of CD4+ target cells.   Since a variant $v$ will have population size of order $L$, for $L$ less than $10000$ the $v$ variant population will have a frequency on the order of $10^{-5}$.  Such low frequencies cannot be sampled by even deep sequencing approaches so ignoring the dynamics does not limit connection to data.  Further, the error produced by ignoring such a low frequency variant in computing birth-death process dynamics is dominated by the error produced by the stochastic-deterministic decomposition, the sampling error associated with the data,  the error produced by solving equations such as (\ref{E:det_analogue}) numerically, and the stochasticity of the birth-death process.   For example, in the context of Figure \ref{F:L_2000} discussed in the previous section, the difference between $T(15)$ and $I_F(15)$ under the \textit{Modified Stochastic Interval Step} as opposed to the \textit{Stochastic Interval Step} is approximately $10^{-7}$.  Such a difference is clearly dominated by the stochasticity of the birth death process and the error of the stochastic-deterministic decomposition as shown in Figures \ref{F:L_2000}A-B.
	
	Using this new simulation approximation, the stochasticity of the birth-death process is reduced to the variables $I_v(t_\spend^{(v)})$ or in other words the draws of the samples $\hat{x}_v$ from the distributions $X_v$.  To see this notice that except for sampling from $\hat{x}_v$ for all variants $v \ne F$, the simulation of the birth-death process is completely deterministic.  We refer to $X_v$ and $\hat{x}_v$ as the pop size distribution and pop size of the $v$ variant population because, intuitively, $\hat{x}_v$ determines how soon the $v$ variant population has significant frequency and hence 'pops up' in the data.   

	For appropriate choices of $\epsilon$ and $L$ the stochasticity of the birth-death process can be further simplified.   For concreteness, consider the escape graph of Figure \ref{F:CH58_escape_graph}.  If $\epsilon$ is chosen large enough and $L$ is chosen small enough then $t_\spend^{(M1)} < t_\spstart^{(M12)}$, i.e. the stochastic interval of $M1$ will end before the stochastic interval of $M12$ starts.   In this case, the stochastic interval of $M1$ will be considered first, the distribution of $X_{M1}$ generated, and $\hat{x}_{M1}$ sampled.   Only once $\hat{x}_{M1}$ has been sampled will the dynamics be run forward to the $M12$ stochastic interval.   Then, $X_{M12}$ will be generated and $\hat{x}_{M12}$ will be sampled.  Putting all this together, $\hat{x}_{M1}$ and $\hat{x}_{M12}$ are states in a Markov chain.   Further, the dynamics of the birth-death process are deterministic other than the choices of $\hat{x}_{M1}$ and $\hat{x}_{M12}$.   In this sense, the birth-death process is reduced to a Markov chain.   For more complex escape graphs, the idea is the same with the linear Markov chain of the example being replaced by a Markov chain on a general escape graph.  The pop size of a vertex $v$ depends on the pop sizes sampled for vertices that have an edge pointing to it and the Markov chain begins from the founder vertex and moves outwards.
	
	In the context of the escape graph in Figure \ref{F:CH58_escape_graph}, forcing the $M1$ stochastic interval to end before the $M12$ stochastic interval begins requires,
\begin{equation}
\mu k T(t_\spend^{(M1)}) I_{M1}(t_\spend^{(M1)}) < \epsilon.
\end{equation}
The product $k T(t)$ is between $1-2$ in early infection and, generally, can be taken near $1$.  Then recalling that $I_{M1}(t_\spend^{(M1)})$ is sampled from $X_{M1}$, we always have $I_{M1}(t_\spend^{(M1)})$ of order $L$.  All this gives the requirement for $\epsilon$, which holds generally, that $\epsilon > \mu L$.  
	
	Choosing an $\epsilon > 0$ means that mutations of variant $v'$ to $v$ occurring prior to the start of $v$'s stochastic interval will be ignored.  Such mutations have a probability of order $\epsilon$ of occurring with the exact value depending on $k$, typically though the probability is slightly less than $\epsilon$.   All this means that with probability $1-\epsilon$, no such mutations occur and our assumption of $\epsilon > 0$ does not affect the simulation of the birth-death process.   Importantly, our posteriors and p-values are accurate for the subset of the birth-death process realizations that do not have such early mutations.

\subsection{Computing the Posterior}

	Let $S(t)$ be the state at time $t$ of the populations tracked by the birth-death process, e.g. for the escape graph of Figure \ref{F:CH58_escape_graph}, $S(t) = (T(t), I_F(t), I_{M1}(t), I_{M12}(t))$.  Let $\hat{\D}(t_1, t_2)$ be frequency data for the variants of the escape graph collected at time points $t_1, t_2$.  We choose two time points simply for concreteness, any number of data time points are possible.  For example, the CH58 data discussed in the Results section includes three time points, see Table \ref{T:CH58_data}.  Letting $\theta$ represent the parameters of the birth-death process for which we want to build a posterior, we can simulate the birth-death process and generate samples for $S(t_1)$, $S(t_2)$.   Let $\D(S(t_1), S(t_2))$ be the data generated by simulating the birth-death process and then simulating the task of sampling sequences.  By this we mean that first, through simulation, $S(t_1)$, $S(t_2)$ must be sampled to establish the exact frequencies of the variants at times $t_1$, $t_2$.  Then, since the data produced in \cite{Goonetilleke_2009_JEM} comes from sampled HIV sequences, hypothetical samples must be drawn at time $t_1$ and $t_2$ to form simulated data.
	
	Given a prior for $\theta$, $\pi(\theta)$, our goal is to compute a posterior of $\theta$ conditioned on the data.  More precisely, we aim to compute
\begin{equation}  \label{E:theta_posterior}
P(\theta \ | \ \D(S(t_1), S(t_2)) = \hat{\D}(t_1, t_2)).
\end{equation}
However, it is easier to compute a posterior for $\theta, S(t_1), S(t_2)$,
\begin{equation}  \label{E:full_posterior}
P(\theta, S(t_1), S(t_2) \ | \ \D(S(t_1), S(t_2)) = \hat{\D}(t_1, t_2)),
\end{equation}
and (\ref{E:theta_posterior}) can be obtained from (\ref{E:full_posterior}) by treating $S(t_1)$ and $S(t_2)$ as nuisance parameters.

Bayesian posteriors are often computed through Markov chain Monte Carlo (MCMC) methods, see chapter $7$ of \cite{Phylo_Book} for a review of MCMC theory applied to viral data and \cite{MCMC_Book} for a general review.  In our context, implementing such an approach depends on being able to compute the probability,
\begin{equation}  \label{E:sample_dist}
P(\theta, S(t_1), S(t_2) \ | \ \D(S(t_1), S(t_2)) = \hat{\D}(t_1, t_2)). 
\end{equation}
Once (\ref{E:sample_dist}) can be computed, various MCMC methods allow one to sample from the posterior of $\theta$.  Specifically, 
we implement a Metropolis-Hastings based MCMC.  Such an approach is not affected if instead of computing (\ref{E:sample_dist}) we compute
\begin{equation}  \label{E:sample_dist_2}
P(\theta, S(t_1), S(t_2),  \D(S(t_1), S(t_2)) = \hat{\D}(t_1, t_2)).
\end{equation}
Indeed, (\ref{E:sample_dist_2}) is identical to (\ref{E:sample_dist}) up to a constant factor and such a proportional factor has no affect on a Metropolis-Hastings MCMC.  

(\ref{E:sample_dist_2}) can be expressed as a product of simpler conditional probabilities,
\begin{align}
P(\theta & , S(t_1), S(t_2), \D(S(t_1), S(t_2)) = \hat{\D}(t_1, t_2)) 
\\ \notag
	& = P(\D(S(t_1), S(t_2)) = \hat{\D}(t_1, t_2) \ | \ S(t_1), S(t_2), \theta)     	P(S(t_1), S(t_2) \ | \  \theta) \pi(\theta).
\end{align}
The factor $P(\D(S(t_1), S(t_2)) = \hat{\D}(t_1, t_2) \ | \ S(t_1), S(t_2), \theta)$ can be interpreted as a sampling probability.  That is, conditioned on knowing $S(t_1), S(t_2)$ and hence the frequencies of the variants at times $t_1, t_2$, what is the probability of drawing the data.   Computation of such probabilities is standard, see for example the methods of \cite{Ganusov_2011_J_Virology}.   However, $P(S(t_1), S(t_2) \ | \ \theta)$, the probability of a given system state at $t_1, t_2$ given a parameter choice, is not standard.

	Our approach is to use the Markov chain reduction to replace $S(t_1), S(t_2)$ by $\hat{x}_v$ for all variants $v \ne F$.  Since the birth-death process under our approximation is completely determined by the pop size samples, the $\hat{x}_v$ are interchangeable with the $S(t)$.   To form the posterior, we replace the $S(t_1), S(t_2)$ values above by the $\hat{x}_v$.   Then the factor $P(S(t_1), S(t_2) \ | \ \theta)$ becomes $P(\hat{x}_v \text{ for } v \ne F \ | \ \theta)$.   This second expression is simply the probability that a Markov chain takes on a certain state.   Since we know the distribution of each $\hat{x}_v$, $P(\hat{x}_v \text{ for } v \ne F \ | \ \theta)$ can be computed in a standard manner.
	
	We use a random walk Metropolis-Hastings algorithm on $\theta$ and the $\hat{x}_v$ to form the posterior.  For instance, to form the posterior for patient CH58, $\theta$ is $6$ dimensional and the $\hat{x}_v$, namely $\hat{x}_{M1}$ and $\hat{x}_{M12}$, are two dimensional.   Our MCMC then operates on an $8$ dimensional state space.   To compute a single step of the MCMC takes approximately $.2$ seconds on an Intel I7-2600 using our C++ implementation.  For all results below we generated $1.5$ million steps for the Markov chain which takes between $2-3$ days.  Various improvements in the C++ implementation, most notably multi-threading, should significantly improve computation times.   Mixing times for the MCMC are roughly on the order of $5,000$ steps.

\subsection{Hypothesis Testing}

	Given an escape graph and a choice of parameters for the birth-death process, our goal is to test the null hypothesis that the data is formed by the model.  Here we let $\theta $ represent all the parameters of the birth-death process and the underlying escape graph, as opposed to the previous section where $\theta$ represented only the parameters included in the posterior.  The challenge lies in computing a p-value.  More precisely, our goal is to compute the p-value of the data, $\hat{\D}(t_1,t_2)$, given $\theta$.
Since the data is multidimensional, the notion of a p-value is not a priori well-defined.  However, as in the case of posterior computations, the Markov chain reduction allows for a simplification.

	Using the escape graph of Figure \ref{F:CH58_escape_graph} as a concrete case, for a given $\theta$ there will be a pair of pop sizes $\hat{x}_{M1}, \hat{x}_{M12}$ for which $P(\D(\hat{x}_{M1}, \hat{x}_{M12}) = \hat{D}(t_1, t_2) \ | \ \theta)$ is maximized, here we are replacing $\D(S(t_1), S(t_2))$ by $\D(\hat{x}_{M1}, \hat{x}_{M12})$ since the pop sizes completely determine the dynamics of the birth-death process.    Generically, the maximum need not be unique, however this is the case for the models considered in the Results section.  (The case of non-unique maximums can be addressed, but we do not explore that issue in this paper.)  Label the $\hat{x}_{M1}, \hat{x}_{M12}$ that achieve the maximum as $x^\text{data}_{M1}, x^\text{data}_{M12}$.   Then we can assess the p-value of $\hat{D}(t_1, t_2)$ by considering where $(x^\text{data}_{M1}, x^\text{data}_{M12})$ falls in the two-dimensional density of $X_{M1}, X_{M12}$.  In other words, we use pop sizes that maximize the likelihood of the data to determine the p-value.
	
	More precisely, let $\Phi_{M1}$ and $\Phi_{M12}$ be the cumulative densities of $X_{M1}$ and $X_{M12}$ respectively.   Then let $\Phi_{M1}(x^\text{data}_{M1}) = q_{M1}$ and $\Phi_{M12}(x^\text{data}_{M12}) = q_{M12}$.   If $x^\text{data}_{M1}$ was chosen from the density of $X_{M1}$, $q_{M1}$ would be uniformly distributed on $[0,1]$.  The same holds for $q_{M12}$.  Further, note that under the null hypothesis $q_{M1}$ and $q_{M12}$ are independent because the sampling from the the pop size distributions $X_{M1}$ and $X_{M12}$ occurs independently even though the distribution $X_{M12}$ depends on $X_{M1}$.  
	
	We can then define the p-value by computing the probability with which two independent, uniform random numbers are more 'extreme' than $q_{M1}, q_{M12}$.   More precisely, let $\Omega$ be the unit square produced by combining all pairs of numbers both falling in $[0,1]$.  Let $f(x)$ for $x \in [0,1]$ be the shortest distance from $x$ to either $0$ or $1$, i.e. $f(x) = \min(x,1-x)$. Then we define the p-value associated with $\theta$ by,
\begin{equation}  \label{E:p_value_def_2}
\text{p-value of } \theta = P(f(u_1)f(u_2) < f(x^\text{data}_{M1})  f(x^\text{data}_{M1}))
\end{equation}
where $u_1, u_2$ are uniform samples from $[0,1]$.
In words, $f(x^\text{data}_{M1})  f(x^\text{data}_{M1})$ is a measure of how far the pair $x^\text{data}_{M1}, x^\text{data}_{M1}$ is from the boundary of $\Omega$ and we set the p-value to be the probability that two uniform samples from $[0,1]$ are closer to the boundary under this measure than $x^\text{data}_{M1}, x^\text{data}_{M1}$.   For example, the pair $.5, .5$ will have a measure of $.5^2$ from the boundary of $\Omega$ which is greater than any other point.  Hence $.5, .5$ would have a p-value of $1$.  In practice, we evaluate (\ref{E:p_value_def_2}) through Monte Carlo methods by sampling $10^7$ pairs of $u_1, u_2$ and determining the fraction of times $f(u_1)f(u_2) < f(x^\text{data}_{M1})  f(x^\text{data}_{M1})$.  For escape graphs with more variants, the same idea is employed but in higher dimension.
	
	This definition of p-value may seem non-intuitive as we are comparing samples from the pop size distributions rather than data values.  However, we aim to quantify the degree to which the data is unusual for the model.   To assess this, we map the data to the probability space of the model through the Markov chain reduction.   Then, since the Markov chain has a simple probabilistic structure, p-values can be defined and computed more readily.

\section{Results}

	The data we consider is summarized in Figure $2$ of \cite{Goonetilleke_2009_JEM}, although we also exploit linkage information provided by the sequence data not reflected in the figure.  In \cite{Goonetilleke_2009_JEM}, the time points of the data are measured in days since patient identification.  So for example, day $0$ represents the time of initial identification rather than the day of initial infection.  To make this distinction clear, we always use 'day $x$' to mean $x$ days since patient identification and '$t=x$' to mean $x$ days since initial infection.
	
	The data we use for CH58 is composed of time points day $9$ and day $45$ at which $7$ and $9$ sequences are available as well as day $0$ data in which the founder variant is homogeneous.   The stochasticity associated with such small sample sizes is large and overwhelms the stochasticity of the birth-death process that we aim to highlight.  For example, in \cite{Ganusov_2011_J_Virology}, the escape rate at epitope ENV EL9 (introduced and discussed below) is estimated as $.1$ but with a $95\%$ confidence interval of $[.005, .808]$ (see Figure S3, row ENV 581 in \cite{Ganusov_2011_J_Virology}).
Since the aim of this paper is to highlight the model, the results presented for CH58 below assume that the frequency data is generated from $1000$ sequences at each timepoint, a modest value for deep sequencing datasets.  Our posteriors are then analogous to the $.1$ value of \cite{Ganusov_2011_J_Virology} with the stochasticity described by the posterior coming mainly from the birth-death process but also from the sampling of $1000$ sequences per timepoint.  For CH40, deep sequencing is available in \cite{Fisher_2010_PLOS_One} and our results correspond to the level of sampling given in that dataset.

	For both patient CH58 and CH40 we choose $L=2000$, making a conservative choice to insure accuracy for the stochastic-deterministic decomposition.   For patient CH40, since the escape graph contains only edges emanating from the founder, we can choose $\epsilon = 0$ without worry of overlapping stochastic intervals.   In this case then, our simulation and the resultant priors and hypothesis tests apply to all realizations of the birth death process.  For patient CH58, we choose $\epsilon = .06$ to insure a separation between the $M1$ and $M12$ stochastic intervals.  As a result we ignore roughly $6\%$ of the birth-death process realizations. 
	
	 For patients CH58 and CH40 we set $k = 2.6 * 10^{-3}$ and $k= 3 * 10^{-3}$, respectively.  For both patients we set $d = .01$, $\lambda = 10^8$.  The parameters were chosen to match the time of peak viral loads suggested by the data as well as to fall within a biologically reasonable range \cite{Stafford_2000_J_Theor_Bio, DeBoer_2007_J_Virology, Perelson_1993_Math_BioSci}.  Choices for the $\delta$ death rates and attack intervals are discussed separately for each patient.
	
\subsection{Patient CH58:}

	For patient CH58, we consider data collected in the first three time points : day $0$, day $9$, and day $45$.   The patient was identified in Fiebig stage II and according to viral load data, day $0$ corresponds to a time slightly prior to peak viral load (see Figure S1 in \cite{Goonetilleke_2009_JEM}).   We make the rough estimate that $t=20$ corresponds to day $0$.   Other choices in the range $t \in [15,25]$ are reasonable, but the basic results we present do not change.  
	
	  In \cite{Goonetilleke_2009_JEM}, two regions on the founder genome  experienced rapid, early escape.   The first region contains the ENV EL9 epitope (see CH58.e in Figure $2$ of \cite{Goonetilleke_2009_JEM}) which elicited a "very early dominant but transient" T cell response.   The second region which we label as ENV 830 corresponding to its location relative to the HXB2 genome (see CH58.g in Figure $2$ of \cite{Goonetilleke_2009_JEM}) was not associated with a known epitope or experimentally identified T cell response.   Three other regions of early escapes were found in \cite{Goonetilleke_2009_JEM}, but the escapes at these regions came slightly after the escapes at ENV EL9 and ENV 830 and were still in their initial stages at day $45$.  ICS data showed a strong memory CD8 T cell response to ENV EL9 at day $21$, but a very weak response by day $45$, see Figure 7 in \cite{Goonetilleke_2009_JEM} and Figure S5B in \cite{Ganusov_2011_J_Virology}.  
	  
	  Table \ref{T:CH58_data} shows escape percentages at ENV EL9 and ENV 830 for the three timepoints.   The first three columns give, from left to right, the percentage of samples that have mutations in ENV EL9 and the ENV 830 region; ENV EL9 only; and ENV 830 region only; respectively.    The fourth column, corresponds to the percentages found in Figure $2$ of \cite{Goonetilleke_2009_JEM} and gives the percentage of mutants in the ENV EL9 region with the status of the ENV 830 region ignored.    Table \ref{T:CH58_data} suggests that the escapes at ENV EL9 and ENV 830 overlap.  Our goal in this section is to investigate the escape rates of the sweeps at these two genome locations in a way that accounts for their interaction.

\begin{table}[h]
\begin{center}
\caption{CH58 Data}
\begin{tabular}{|c||c|c|c||c|}  
\hline 
day  & ENV EL9 and 830 & ENV EL9 only & ENV 830 only & ENV EL9 ignoring 830\\
\hline
$0$  & $0\%$ &  $0\%$ & $0\%$ & $0\%$
\\
\hline
$9$  & $0\%$ &  $28\%$ & $0\%$ & $28\%$
\\
\hline
$45$  & $89\%$ &  $0\%$ & $0\%$ & $89\%$
\\
\hline
\end{tabular}
\label{T:CH58_data}
\end{center}
\end{table}

	The data presented in the previous two paragraphs suggests an escape graph of the form given in \ref{F:CH58_escape_graph} with $F$, $M1$, and $M12$ playing the role of variants of founder type, variants with escape mutations in ENV EL9 alone, and variants with mutations in ENV EL9 and the ENV 830 region.  We separate the dynamics into three attack intervals, $t=0$ to $t=15$, $t=15$ to $t=30$, and $t = 30$ to $t=65$, through which we form piecewise constant estimates for $\delta_F(t)$, $\delta_{M1}(t)$ and $\delta_{M12}(t)$.  The time interval $t=0$ to $t=15$ is meant to model the period prior to immune system attack.  The split of the second and third attack intervals at $t=30$ allows for the change in T cell attack at ENV EL9 implied by the data.   
	
	Table \ref{T:CH58_parameters} gives the $\delta$ parameters of the model, $9$ in all.  However, $\delta_{M12,1}$ (the death rate during $t \in [0,15]$ of M12 variants) is not relevant because there are no M12 variants during that time period.   We set $\delta_{F,1} = .4$ following standard estimates on HIV infected cell lifetimes.  The data does not bound the value of $\delta_{M1,3}$ from above.  Indeed, there are no $M1$ variants in the data at day $45$ so there can be no upper bound on the death rate of $M1$ variants during the interval between day $9$ and day $45$.  Consequently, we do not view $\delta_{M1,3}$ as adding a full dimension to the posterior.   Finally then, the posterior is $6$ dimensional corresponding to the parameters : $\delta_{M1, 1}$, $\delta_{F,2}$, $\delta_{M1,2}$, $\delta_{M12,2}$,  $\delta_{F,3}, \delta_{M12,3}$.

\begin{table}[h]
\begin{center}
\caption{CH58 Parameters}
\begin{tabular}{|c||c|c|c|}  
\hline 
variant  & $\delta$ during $[0,15]$ & $\delta$ during $[15,30]$ &$\delta$ during $[15,65]$ \\
\hline
F  & $\delta_{F,1}$ &  $\delta_{F,2}$ & $\delta_{F,3}$ 
\\
\hline
M1  & $\delta_{M1,1}$ &  $\delta_{M1,2}$ & $\delta_{M1,3}$ 
\\
\hline
M12   & $\delta_{M12,1}$ &  $\delta_{M12,2}$ & $\delta_{M12,3}$ 
\\
\hline
\end{tabular}
\label{T:CH58_parameters}
\end{center}
\end{table}

In terms of our parameters, we define
\begin{flushleft}
escape rate $1$ = $\delta_{F,2} - \delta_{M1,2}$\\
escape rate $2$ = $\delta_{F,3} - \delta_{M12,3}$
\end{flushleft}
which quantify the rate of escape at the two regions.  
This matches, in terms of our model, the escape rate definition given in \cite{Asquith_PLOS_Biology_2006} and used in \cite{Goonetilleke_2009_JEM} and \cite{Ganusov_2011_J_Virology}.

\begin{figure} [h]
\begin{center} 
\includegraphics[width=4in]{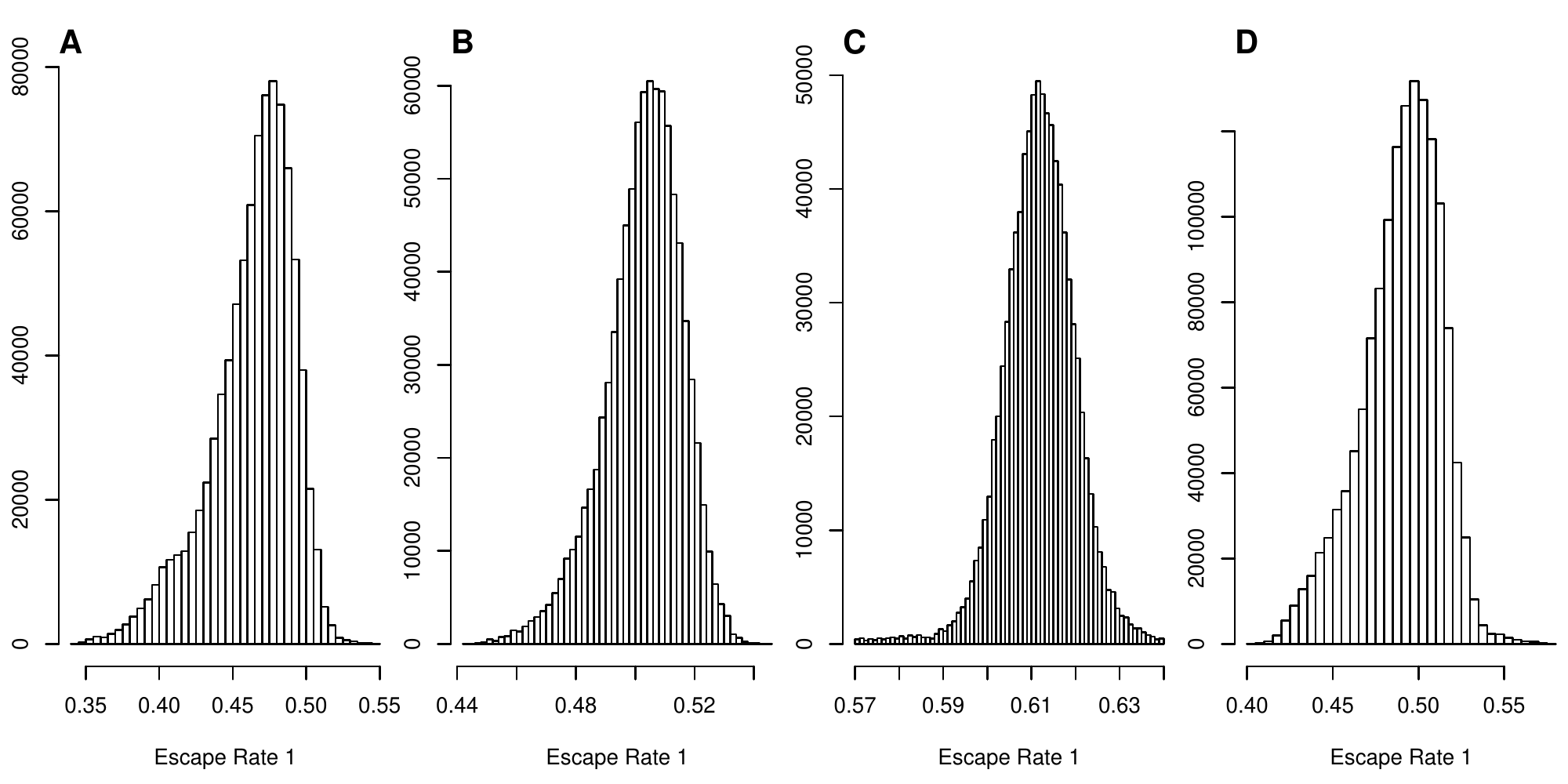}
\end{center}
\caption{\textbf{Escape Rate $1$ Posteriors}.  Escape rate $1$ measures the rate of escape at epitope ENV EL9.  Posteriors shown in A-C were generated with a uniform prior on $[0,2]$ for all parameters in Table \ref{T:CH58_parameters}, except that $\delta_{F,1} = .4$ and (A) $\delta_{M1,1} = .4$ (B) $\delta_{M1,1} = .6$ (C) $\delta_{M1,1} = 10$.  The posterior in D was generated using a uniform prior of $[.4,2]$ for all parameters, except $\delta_{F,1} = .4$ and  $\delta_{M1,1}$ was assigned a uniform prior on $[.4,.7]$.  All posteriors were generated by running an MCMC for $1.5 * 10^6$ steps.} 
\label{F:escape_rates_1}
\end{figure}

	Figures \ref{F:escape_rates_1}A-C give escape rate $1$ posteriors with $\delta_{M1,1} = .4, .6, 10$, respectively.  For all other parameters our prior was a uniform distribution on $[0,2]$.   The value $\delta_{M1,1} = 10$ is not plausible biologically, but is useful in understanding the relationship between $\delta_{M1,1}$ and escape rate $1$.     As $\delta_{M1,1}$ rises, the number of ENV EL9 mutants in existence at $t=15$, the time immune response begins, drops.  Biologically, a larger $\delta_{M1,1}$ means that the ENV EL9 mutants are less fit.  With a lower number of mutants at $t=15$, the escape rate on $[15,30]$ needs to be higher in order to fit the data, specifically the day $9$ frequencies.  Since an ENV EL9 mutant is not expected to be more fit than the founder variant, i.e. $\delta_{M1,1} > .4$, and since $\delta_{M1,1} < 10$ is reasonable, Figures \ref{F:escape_rates_1}A-C show that the posterior of escape rate $1$ falls somewhere in the range of $.35$ to $.6$.  Figure \ref{F:escape_rates_1}D gives the posterior of escape rate $1$ when using a uniform prior of $[.4, .7]$ on $\delta_{M1,1}$ and $[.4,2]$ on all other parameters.  This prior assumes a fitness cost for all mutants over the founder variant.  As can be seen, Figure \ref{F:escape_rates_1}D is roughly a combination of Figures \ref{F:escape_rates_1}A and \ref{F:escape_rates_1}B and gives a range for escape rate $1$ of $.4$ to $.55$.

\begin{figure} [h]
\begin{center} 
\includegraphics[width=4in]{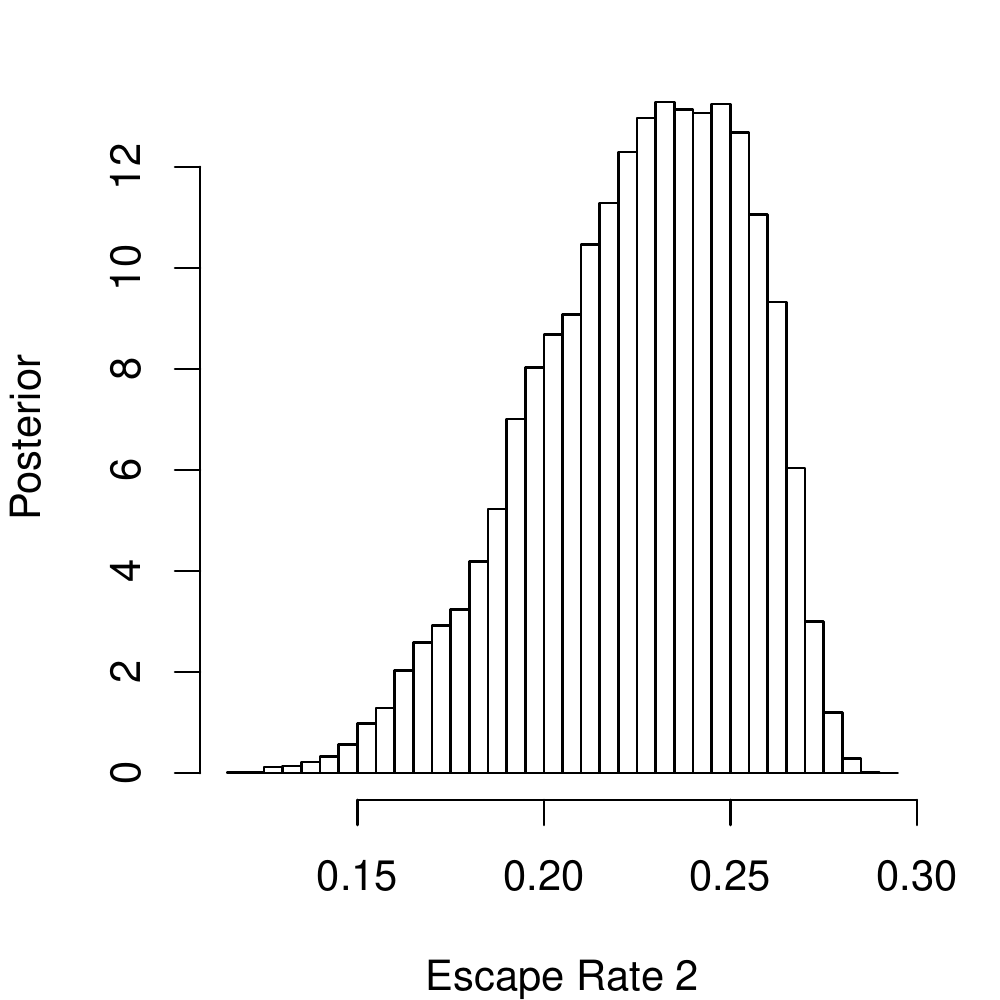}
\end{center}
\caption{\textbf{Escape Rate $2$ Posterior}.  Escape rate $2$ measures the rate of escape in the region including ENV 830.  The posterior was generated with a uniform prior on $[0,2]$ for all parameters in Table \ref{T:CH58_parameters} except $\delta_{F,1} = .4$ and  $\delta_{M1,1}, \delta_{M1,2}$ were assigned uniform priors on $[.4, .7]$ and $[.4, 1]$, respectively.  See Figure \ref{F:escape_rates_1} for all other parameter values.  The posterior was generated by running an MCMC for $1.5 * 10^6$ steps.} 
\label{F:escape_rates_2}
\end{figure}

\begin{figure} [h]
\begin{center} 
\includegraphics[width=4in]{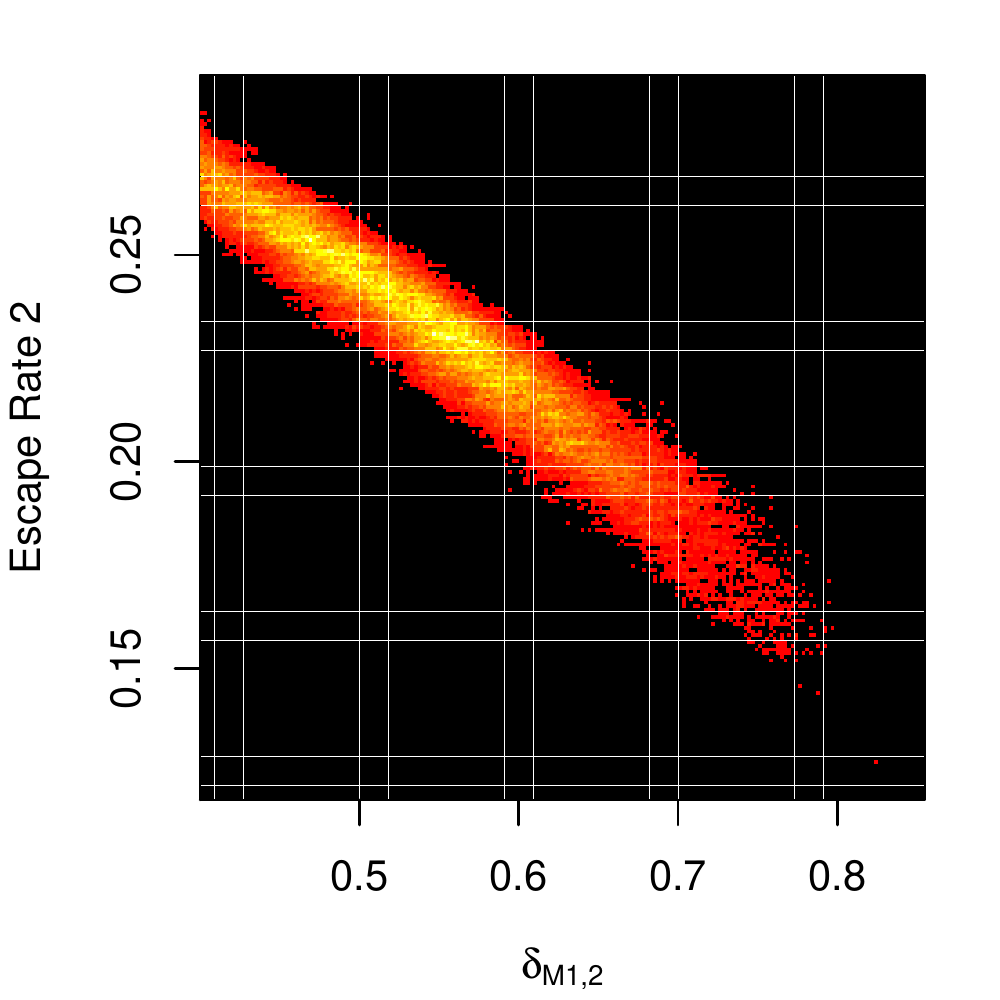}
\end{center}
\caption{\textbf{Posterior of escape rate $2$ and $\delta_{M1,2}$}.  See figure \ref{F:escape_rates_2} for priors and \ref{F:escape_rates_1} for all other parameter values.} 
\label{F:heat_map}
\end{figure}

	Figure \ref{F:escape_rates_2} gives the posterior for escape rate $2$ which falls roughly in the range $[.1, .3]$.   The posterior was generated with a prior on $\delta_{M1,1}$ and $\delta_{M1,2}$ that is uniform on $[.4, .7]$ and $[.4, 1]$, respectively.  All other parameters had a uniform prior on $[0,2]$.    Altering $\delta_{M1,2}$ has an analogous effect on escape rate $2$ as altering $\delta_{M1,1}$ has on escape rate $1$.   A relatively large value for $\delta_{M1,2}$ means that ENV EL9 mutants will die off more quickly, allowing ENV EL9 + ENV 830 mutants to rise more quickly, and in turn reducing escape rate $2$.  Figure \ref{F:heat_map} gives a heat map for posterior values of the variables $\delta_{M1,2}$ and escape rate $2$ showing this inverse relationship.  
	
	Table \ref{T:parameter_set_a} gives a set of parameter values which we label as parameter set $a$.  We also consider parameter set $b$, which we take identical to set $a$ except with $\delta_{F,2} = 1.05$ rather than $\delta_{F,2} = 1.1$.  Escape rate $1$ for parameter sets $a$ and $b$ is $.5$ and $.45$ respectively, both parameters sets have an escape rate $2$ of $.2$.  Table \ref{T:parameter_set_p_values} gives the pop sizes and p-values for these two parameter sets.  The pop sizes shown in the table are those associated with the p-value, i.e. the pop sizes that maximize the likelihood of the data.    Connecting to Figure \ref{F:escape_rates_1}B (since $\delta_{M1,1} = .6$ in both sets) and Figure \ref{F:escape_rates_2}, parameter set $a$ reflects escape rates in the middle of the posterior while parameter set $b$ has an escape rate $1$ on the left edge of the posterior.   Correspondingly, the overall p-value drops from $.42$ for set $a$ to $.02$ for set $b$.

\begin{table}[h]
\begin{center}
\caption{Parameter Set $a$}
\begin{tabular}{|c||c|c|c|}  
\hline 
variant  & $\delta$ during $[0,15]$ & $\delta$ during $[15,30]$ &$\delta$ during $[15,65]$ \\
\hline
F  & $\delta_{F,1} = .4$ &  $\delta_{F,2} = 1.1$ & $\delta_{F,3} = .9$ 
\\
\hline
M1  & $\delta_{M1,1} = .6$ &  $\delta_{M1,2} = .6$ & $\delta_{M1,3} = 1$ 
\\
\hline
M12   & $-$ &  $\delta_{M12,2} = .4$ & $\delta_{M12,3} = .7$ 
\\
\hline
\end{tabular}
\label{T:parameter_set_a}
\end{center}
\end{table}

\begin{figure} [h]
\begin{center} 
\includegraphics[width=4in]{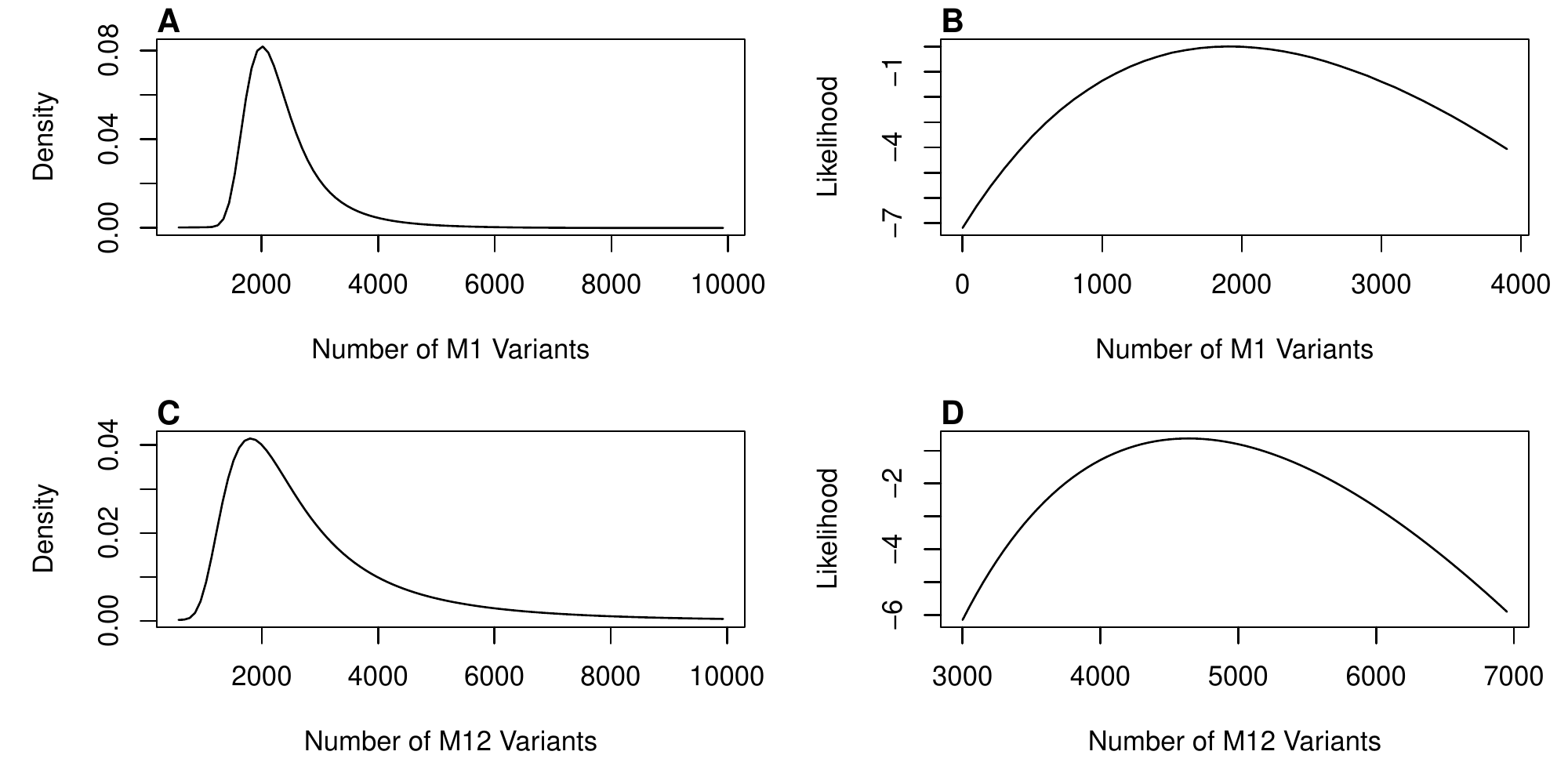}
\end{center}
\caption{\textbf{Parameter Set $a$ Pop Size Distribution and Likelihood}. A-D are generated using parameter set $a$ with  $k = 2.6 * 10^{-3}$ , $d = .01$, $\lambda = 10^8$.  (A) Distribution of $X_{M1}$, the number of $M1$ variants at the end of the $M1$ stochastic interval.  (B) The likelihood of the data given different values for $\hat{x}_{M1}$.  $\hat{x}_{M12}$ is chosen to maximize the likelihood given a value of $\hat{x}_{M1}$.   (C) Distribution of $X_{M12}$, the number of $M12$ variants at the end of the $M12$ stochastic interval.   (D) The likelihood of the data given different values for $\hat{x}_{M12}$ with $\hat{x}_{M1} = 1930$.}
\label{F:parameter_set_a}
\end{figure}

\begin{figure} [h]
\begin{center} 
\includegraphics[width=4in]{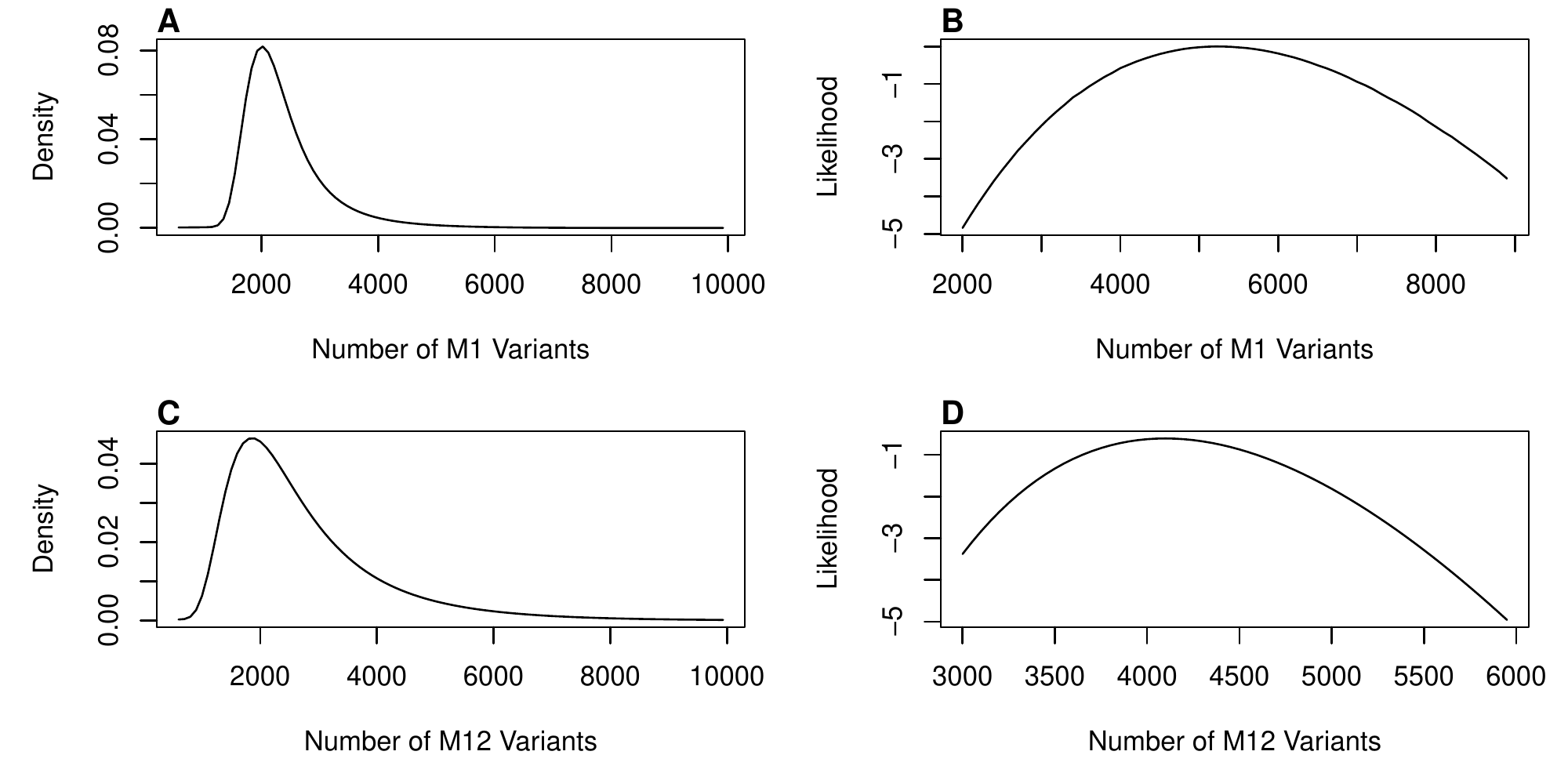}
\end{center}
\caption{\textbf{Parameter Set $b$ Pop Size Distribution and Likelihood}.  A-D are generated using parameter set $b$ with  $k = 2.6 * 10^{-3}$ , $d = .01$, $\lambda = 10^8$.  (A) Distribution of $X_{M1}$, the number of $M1$ variants at the end of the $M1$ stochastic interval.  (B) The likelihood of the data given different values for $\hat{x}_{M1}$.  $\hat{x}_{M12}$ is chosen to maximize the likelihood given a value of $\hat{x}_{M1}$.   (C) Distribution of $X_{M12}$, the number of $M12$ variants at the end of the $M12$ stochastic interval.   (D) The likelihood of the data given different values for $\hat{x}_{M12}$ with $\hat{x}_{M1} = 5330$.} 
\label{F:parameter_set_b}
\end{figure}

	Intuitively, when the birth-death process is defined by parameter set $b$, a rare event has to occur to generate the CH58 data.  On the other hand, the CH58 data is 'typical' for the birth-death process defined by parameter set $a$.  The nature of such rare or typical events are understood through the pop size distribution used to define the p-values.  Focusing on parameter set $a$ for a moment, Figure \ref{F:parameter_set_a}A shows the $M1$ pop size distribution, $X_{M1}$.   More explicitly, the stochastic interval of $M1$ variants is $[6.9, 14.2]$ and so the pop size distribution is precisely the distribution of $I_{M1}(14.2)$.  Figure \ref{F:parameter_set_a}B shows the likelihood of the data given different values for the $M1$ pop size, $\hat{x}_{M1}$.   The maximum likelihood is achieved at $\hat{x}_{M1} = 1930$.   Returning to Figure \ref{F:parameter_set_a}A, a pop size of $1930$ has a p-value of $.7$ as noted in Table \ref{T:parameter_set_p_values}.  Figures \ref{F:parameter_set_a}C,D shows the same graphs but for the $M12$ pop size with the $M1$ pop size fixed at $1930$.   Figure \ref{F:parameter_set_b} shows the same graphs as Figure \ref{F:parameter_set_a} but for parameter set $b$.   Notice from Figure \ref{F:parameter_set_b}B that in this case the maximum likelihood of the $M1$ pop size occurs at $5330$ which is far to the right on the distribution of $M1$ pop size as shown in Figure \ref{F:parameter_set_b}A.   This results in the p-value of $.006$ noted in Table \ref{T:parameter_set_p_values}.    Combining the p-values from the $M1$ and $M12$ pop sizes gives the overall p-value. 
	
\begin{table}[h]
\begin{center}
\caption{Pop Sizes and p-Values for Parameter Sets $a$ and $b$}
\begin{tabular}{|c||c|c|||c|c|c|}  
\hline 
  &  $\hat{x}_{M1}$ & $\hat{x}_{M12}$
	& $\hat{x}_{M1}$ p-value & $\hat{x}_{M12}$ p-value & overall p-value      \\
\hline
$a$  &  $1930$ & $4570$ & $.7$ & $.12$  & $.42$ \\
\hline
$b$  &  $5330$ & $4150$ & $.006$ & $.12$ & $.02$ 
\\
\hline
\end{tabular}
\label{T:parameter_set_p_values}
\end{center}
\end{table}

\begin{figure} [h]
\begin{center} 
\includegraphics[width=4in]{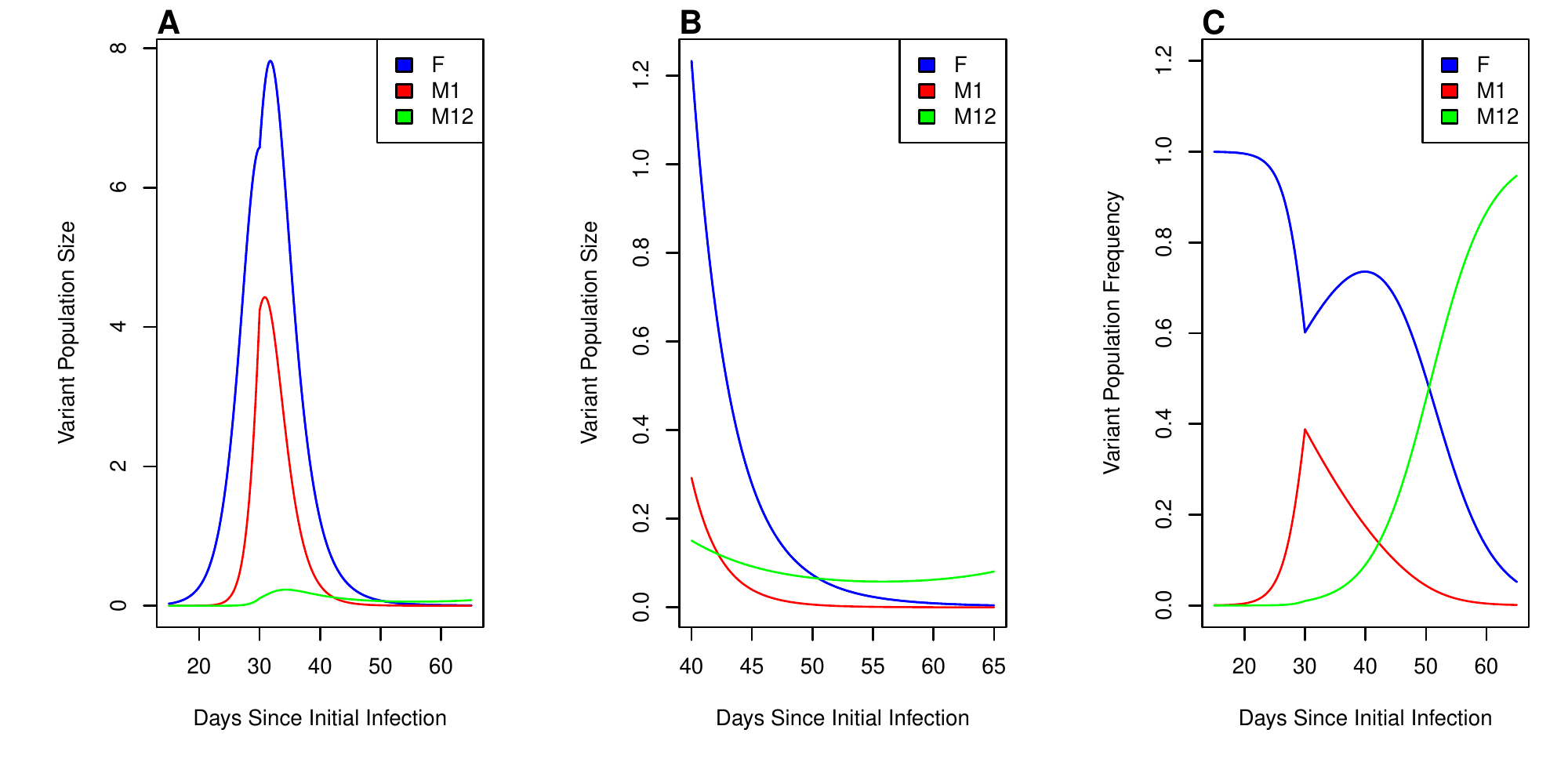}
\end{center}
\caption{\textbf{Dynamics under Parameter Set $a$}.  Dynamics shown are formed using parameter set $a$ with $\hat{x}_{M1}$ and $\hat{x}_{M12}$ given in Table \ref{T:parameter_set_p_values}. (A),(B) population sizes of the three variants in units of $10^9$ infected cells.  (C) population frequencies of the three variants.} 
\label{F:optimum_dynamics}
\end{figure}

	Figures \ref{F:optimum_dynamics}A-C show the dynamics produced by parameter set $a$ with pop size values as given in Table \ref{T:parameter_set_p_values}.  Figures \ref{F:optimum_dynamics}A,B give population sizes in units of $10^9$ infected cells while Figure \ref{F:optimum_dynamics}C gives population frequencies.   There is about a three log difference shown in graph Figure \ref{F:optimum_dynamics}A,B between the time of peak viral load at $t=33$ and $t=65$, this is similar to the change in viral load seen in patient CH58.  The variant percentages shown in Figure \ref{F:optimum_dynamics}C at days $9$ and $45$ are within $1\%$ of those given in Table \ref{T:CH58_data}.

Figure \ref{F:optimum_dynamics}C shows that the escape dynamics associated with parameter set $a$ can be decomposed into two sweeps, one by $M1$, i.e. ENV EL9 mutants, and one by $M12$, i.e. ENV EL9 + ENV 830 mutants.  As can be seen from the graph the ENV EL9 + ENV 830 mutant sweep does not permit the ENV EL9 sweep to finish.   We can give an ad-hoc biological explanation for the values of parameter set $a$. The ENV EL9 mutation comes at an absolute fitness cost of $.2$, i.e. the difference between $\delta_{M1,1}$ and $\delta_{F,1}$.    From $t=15$ to $t=30$, CTL attack at ENV EL9 is enough to make up for this fitness cost and ENV EL9 mutants begin to push out founder variants.   From $t=30$ to $t=65$ the force of CTL attack at ENV EL9 drops, as suggested by the CTL data and attack at other epitopes begins.   Since the CTL attack is no longer focused on ENV EL9, the fitness cost of ENV EL9 mutants causes them to be selected against with respect to founder variants,  However, during this period ENV EL9 + ENV 830 mutants rise up.   ENV 830 region mutations are possibly  compensatory mutation, and so ENV EL9 + ENV 830 mutants are not exposed to the remaining CTL attack at ENV EL9 and do not pay the fitness price of ENV EL9 mutants.  As a result, ENV EL9 + ENV 830 mutants begin to sweep through the population.    We emphasize that this explanation is simply meant to give some intuition to the parameter values and the resulting dynamics,  we do not suggest that our results provide conclusive evidence for this biological interpretation.    

\begin{figure} [h]
\begin{center} 
\includegraphics[width=4in]{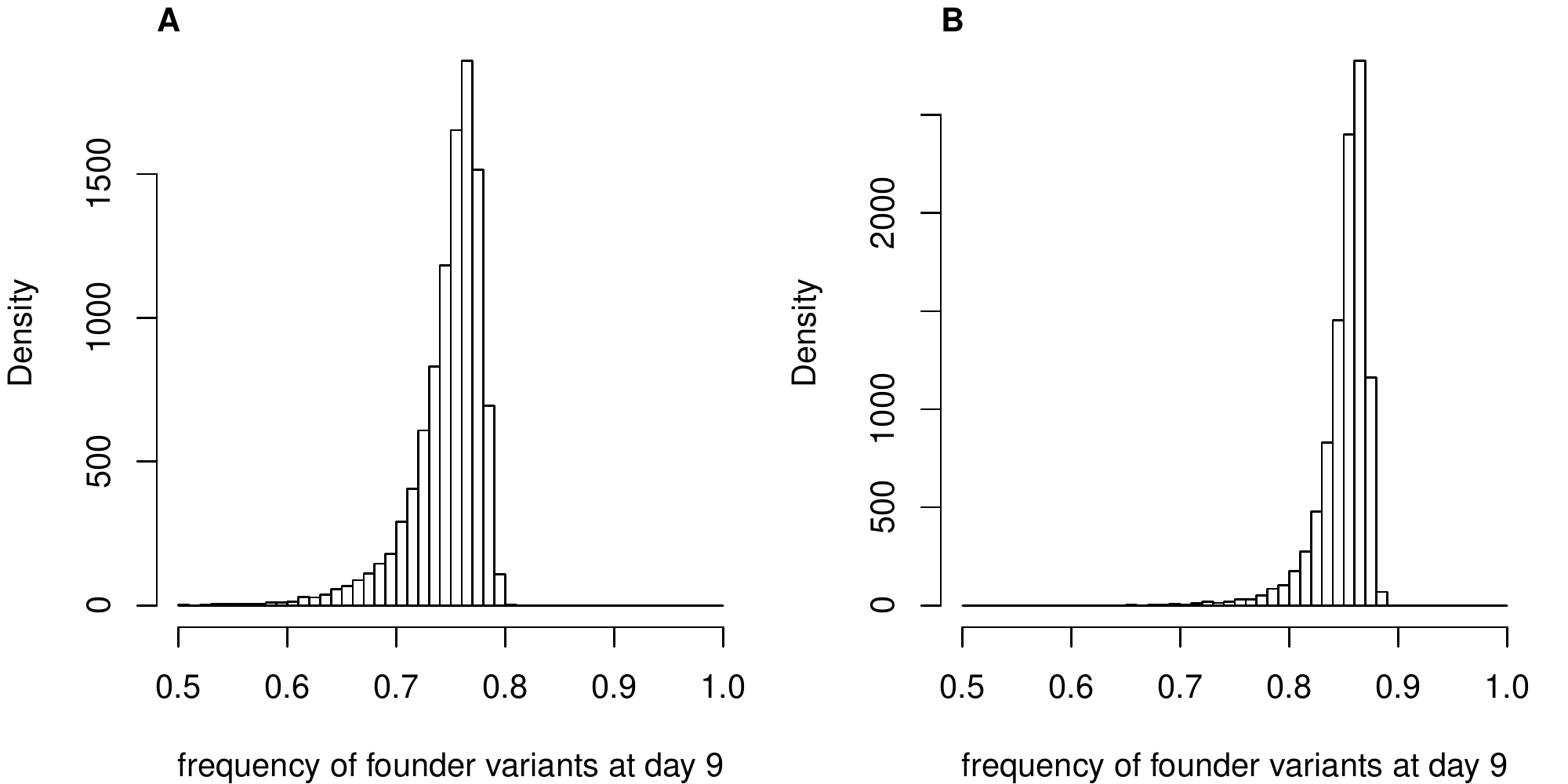}
\end{center}
\caption{\textbf{Founder Variant Frequency Distribution at day $9$}.   (A) parameter set $a$ (B) parameter set $b$.} 
\label{F:stochastic_dynamics}
\end{figure} 

	Finally, we note that when the birth-death process is run under parameter set $a$ without conditioning on the data the dynamics will rarely fit the data.   Figure \ref{F:stochastic_dynamics}A shows the distribution of the frequency of founder variants at day $9$.  CH58 data has the founder variant frequency of $72\%$ at day $9$.  This value indeed has a significant density in  the distribution shown, hence the high p-value of parameter set $a$, but the birth-death process will typically not generate a frequency of $72\%$.    Parameter set $b$ produces a similar density, seen in Figure \ref{F:stochastic_dynamics}B, but with $.72$ located in the tail of the density corresponding to the low p-value of parameter set $b$.

\subsection{Patient CH40:}	 

	For patient CH40, early escapes were identified at two epitopes : NEF SR9 (CH40.t in Figure 2 in \cite{Goonetilleke_2009_JEM}) and GAG CR9 (CH40.b in \cite{Goonetilleke_2009_JEM}).    The first three timepoints sampled are day $0$, $16$, $45$.   NEF SR9 mutants are present in the data by day $16$ while GAG CR9 mutants are not present until day $45$.  With this in mind, we restrict our attention to NEF SR9 and the days $0$ and $16$ allowing us to exploit the deep sequencing data for NEF SR9 described in \cite{Fisher_2010_PLOS_One}.     At day $0$, the data in \cite{Fisher_2010_PLOS_One} is essentially homogeneous for NEF SR9.  Table \ref{T:CH40_data} gives the different NEF SR9 mutants found at day $16$ and is produced from data in Figure $5$ of \cite{Fisher_2010_PLOS_One}.  More precisely, we considered all variants in the figure with frequency greater than $1\%$ at day $16$.  For those variants, frequencies were rounded to the nearest tenth and rescaled (by a factor of $1.033$)  so that frequencies summed to $100\%$.     Each NEF SR9 mutant is associated with a one letter label as specified in the table.

\begin{table}[h]
\begin{center}
\caption{CH40 Data at day $16$}
\begin{tabular}{|c||c|c|}  
\hline 
label  & a.a. sequence & frequency at day $16$     \\
\hline
$F$  & SLAFRHVAR &  50\%   \\
\hline
$Q$ & --------Q & 4.7\% \\ \hline
$H$ & ----H---- & 24.9\% \\ \hline
$N$ & N-------- & 1.9\% \\ \hline
$R$ & R-------- & 3.8\% \\ \hline
$M$ & ------M-- & 5.6\% \\ \hline
$E$ & ------E-- & 2.0\% \\ \hline
$C$ & ----C---- & 5.7\% \\ \hline
$I$ & I-------- & 1.4\% \\ \hline
\end{tabular}
\label{T:CH40_data}
\end{center}
\end{table}

	The striking feature of Table \ref{T:CH40_data} is the high frequency of $H$ variants with respect to the other mutant variants.   This deviation may be explained by differences in fitness, mutation rate, pMHC binding, or CTL recognition.   But before turning to such explanations, it is valuable to consider whether the deviation could be caused by stochastic effects alone.   
	
	To address this issue, we consider different null hypotheses assuming all mutant variants to be identical, meaning under our model that their birth, death, and mutation rates are equal.   
Figure \ref{F:CH40_escape_graph} shows the escape graph we assume.  We separate the dynamics into two attack intervals, $t=0$ to $t=t_A$, $t=t_A$ to $t=36$ (day $16$ corresponds to $t=36$).  $t_A$ is the time at which we assume CTL attack on NEF SR9 to begin and we consider $t_A$ between $t=12$ and $t=19$.  Table \ref{T:CH40_parameters} describes the death rates assumed by the null hypotheses.   All variants, including the founder, have $\delta = .4$ for $t \in [0,t_A]$.  For $t \in [t_A, 36]$,  all mutants share the same death rate denoted $\delta_{M}$ while the founder death rate is denoted $\delta_{F}$.  In total we have three parameters that identify different null hypotheses, $t_A, \delta_F, \delta_M$.  But all null hypotheses parametrized in this manner assume identical mutants.

\begin{table}[h]
\begin{center}
\caption{CH40 Parameters}
\begin{tabular}{|c||c|c|}  
\hline 
variant  & $\delta$ during $[0,t_A]$ & $\delta$ during $[t_A,36]$  \\
\hline
F   & $.4$ &  $\delta_F$ 
\\
\hline
mutants  & $.4$ &  $\delta_M$ 
\\
\hline
\end{tabular}
\label{T:CH40_parameters}
\end{center}
\end{table}

\begin{figure} [h]
\begin{center} 
\includegraphics[width=4in]{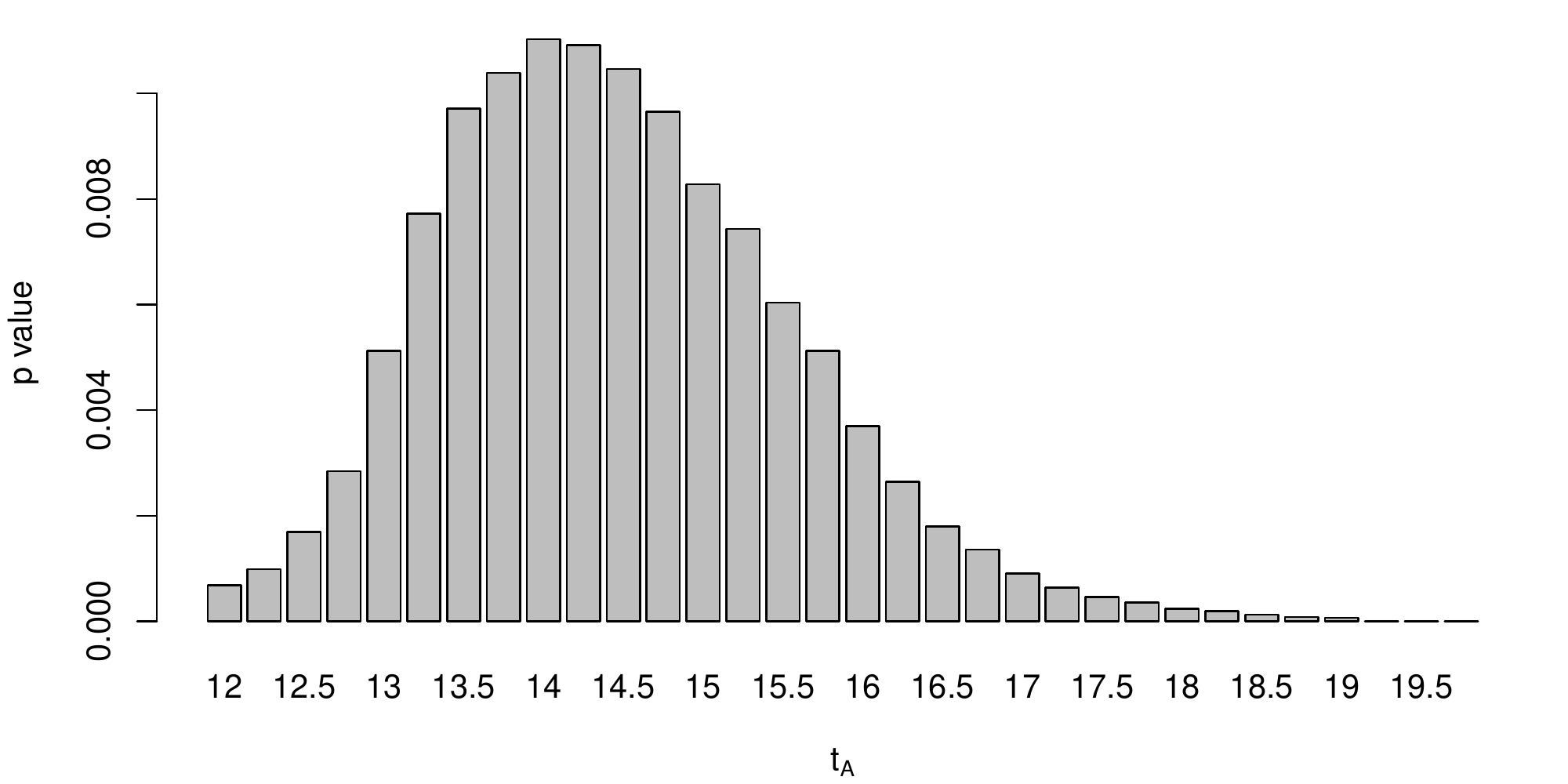}
\end{center}
\caption{\textbf{Null Hypotheses assuming all mutant variants are identical}.  $t_A$ models the time of immune system response in days since infection.} 
\label{F:null_hypothesis_1}
\end{figure}

\begin{figure} [h]
\begin{center} 
\includegraphics[width=4in]{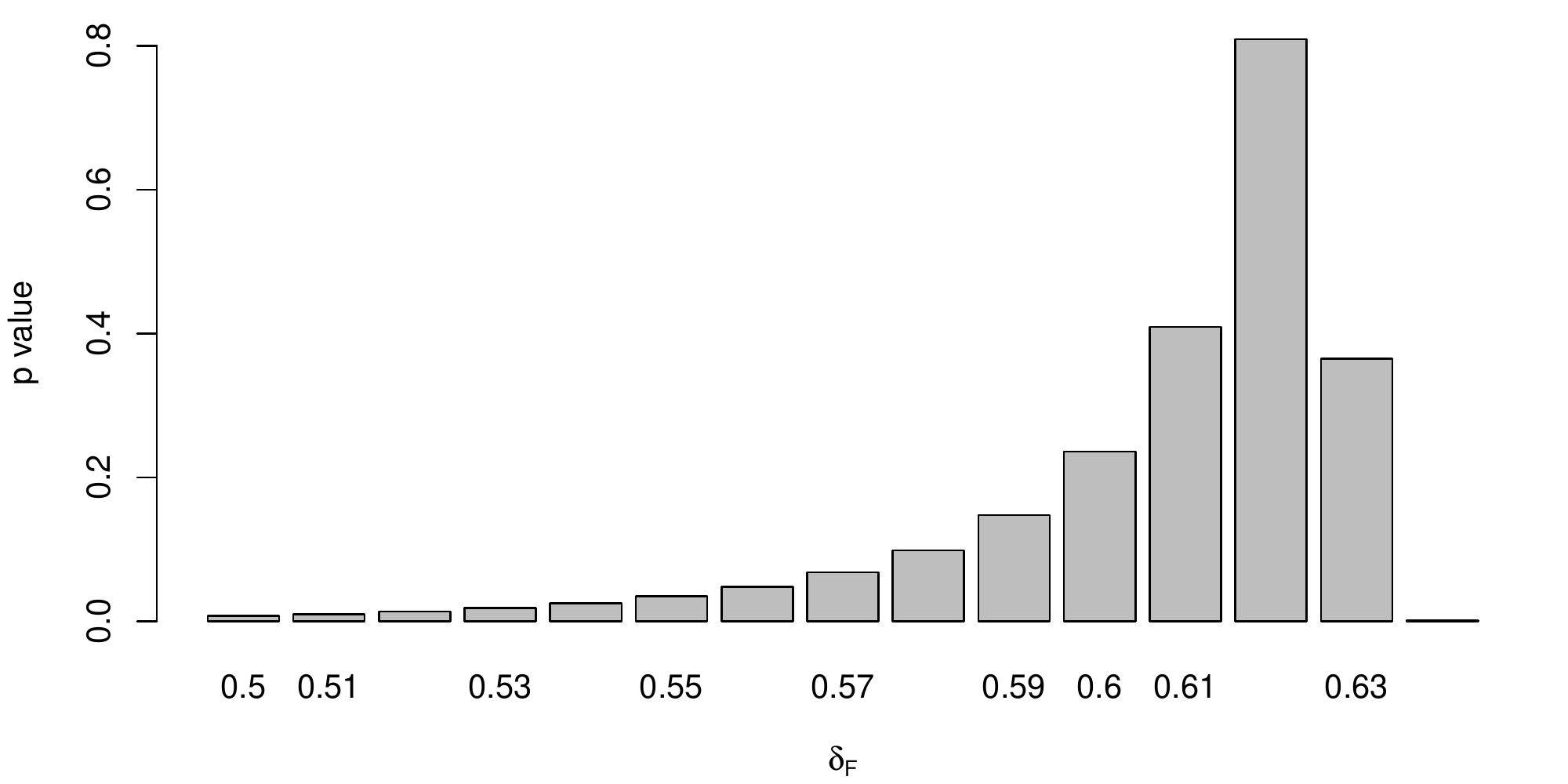}
\end{center}
\caption{\textbf{Null Hypotheses assuming a single mutant variant}.  $t_A$ models the time of immune system response in days since infection.} 
\label{F:null_hypothesis_one_mutant}
\end{figure}

	Figure \ref{F:null_hypothesis_1} provides the p-value for different values of $t_A$ with $\delta_M = .4$, $\delta_F = .56$.    The results are similar in trend with respect to $t_A$ for all $\delta_M, \delta_F$, but with lower p-values.  The maximum p value of $.011$ occurs at $t_A=14$.  The p-value is small, although not exceptionally extreme.   Nevertheless, one might reject this null hypothesis, especially since values of $t_A \ne 14$ generate p-values much lower than $.01$.  Before continuing, we point out that the low p-value is a result of the many variants revealed by deep sequencing.   For example, if all the mutants are grouped into a single mutant with a mutation rate of $8\mu$ and the parameters set at $t_A=15$, $\delta_M = .4$ then Figure \ref{F:null_hypothesis_one_mutant} shows the p-value as we consider hypotheses with varying $\delta_F$.   At $\delta_F = .62$ a p-value of $.81$ is obtained.

\begin{figure} [h]
\begin{center} 
\includegraphics[width=4in]{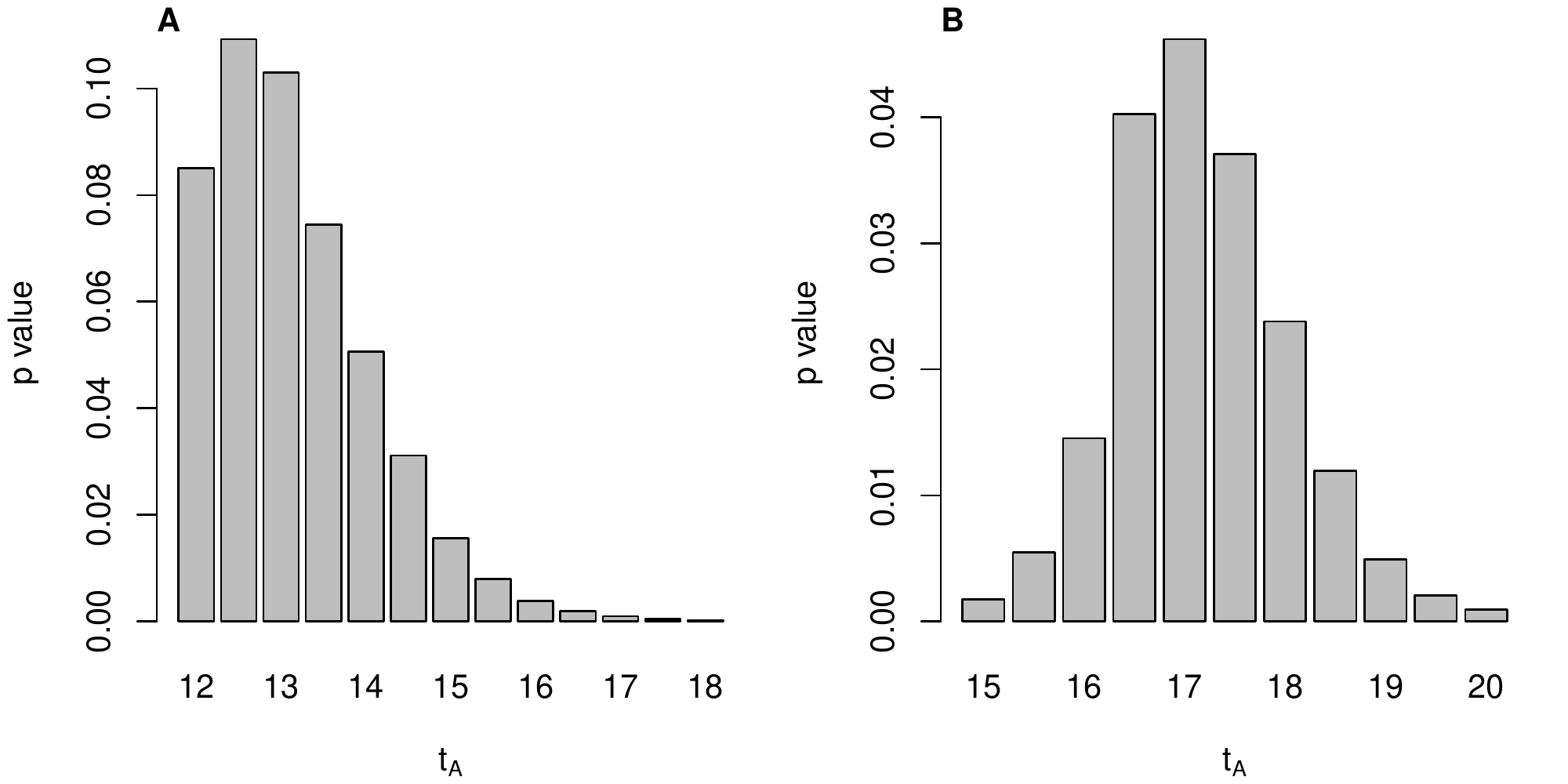}
\end{center}
\caption{\textbf{Null Hypotheses assuming all mutant variants are identical except for variant $H$}.  $t_A$ models the time of immune system response in days since infection.} 
\label{F:null_hypothesis_2}
\end{figure}

	Having rejected the null hypothesis that all mutant variants are identical, a natural second hypothesis would be that all $7$ mutant variants other than $H$ are identical.   We can test this hypothesis using escape graph \ref{F:CH40_escape_graph} and the parameters of Table \ref{T:CH40_parameters}, except that we add a parameter $\delta_H$ as the death rate of $H$ mutants during $[t_A, 36]$.   Figure \ref{F:null_hypothesis_2}A shows p-values under $\delta_H = .4$, $\delta_M = .52$, $\delta_F = .67$ for hypotheses with different $t_A$.   The maximum p-value is $.11$ achieved as $t=12.5$.   Figure \ref{F:null_hypothesis_2}B is generated with identical parameters, except that $\delta_F = .7$.   The maximum p-value is $.05$ which is achieved at the relatively late time of $t_A=17$.   Figures \ref{F:null_hypothesis_2}A,B reflect a general trend, as the death rates on the founder and non-H mutant variants are raised, the attack time must be increased to provide significant p-values.  Roughly, raising death rates provides a greater advantage to $H$ variants that can be offset by shortening the duration of attack.

	While the preceding discussion provides some statistical perspective on early escape in patient CH40, we hasten to add that day $45$ data may well change the picture.  Indeed, a failure to reject the null hypothesis is not proof of its truth.  At day $45$, $Q$ variants come to dominate the population while $H$ variants drop to low levels (see \cite{Fisher_2010_PLOS_One} for more details).   One explanation for these dynamics is that $Q$ variants are more fit than $H$ variants but are exposed to stronger selection by CTLs targeting NEF SR9, similar in some ways to the dynamics seen in patient CH58.  Given these observations, it may be appropriate to remove the assumption in our null hypotheses that all mutant variants are equally fit prior to CTL response.   However, one must also consider the CTL response at GAG and the lack of data linking GAG to NEF for CH40 complicates the analysis.

\section{Discussion}

	HIV escape from CTL attack during acute infection is a remarkable example of evolution.   In the time span of approximately $2$ months the infecting HIV population is targeted by CTL response at $1-3$ epitopes, experiences massive changes in population size, and escapes CTL selective pressure at the targeted epitopes through multiple mutation pathways \cite{Cohen_2011_NJM, Goonetilleke_2009_JEM, Fernandex_J_Virol_2005, Fisher_2010_PLOS_One}.   The resulting HIV selective sweeps are complex in two ways.

	First, HIV selective sweeps are high-dimensional.  As the data collected in \cite{Goonetilleke_2009_JEM} suggests, many variants are involved in an HIV selective sweep and each such variant represents a potential dimension in a model system.   Second, HIV selective sweeps are likely stochastic.   Escape variants must initially have small population sizes which are affected by the stochasticity of life cycle and mutation events. This early stochasticity of escape variants can substantially affect HIV selective sweeps as demonstrated by our model and results.   Indeed, Figure \ref{F:stochastic_dynamics} shows that significant variation in founder frequency soon after peak viral load can be produced by the early stochasticity of escape variant dynamics.

	Corresponding to the complexity of HIV selective sweeps, recent studies have provided high-dimensional datasets that track multiple HIV variants through multiple time points, e.g. \cite{Goonetilleke_2009_JEM, Fernandex_J_Virol_2005, Fisher_2010_PLOS_One}.  In this work we have presented a model of HIV dynamics during acute infection that can incorporate multiple escape variants, through the escape graph, in a stochastic setting, through the birth-death process.   Under this model, we have developed inference methods that allow us to exploit high-dimensional datasets.

	The results presented for patient CH58 highlight the need to consider multiple variants at once.  The early escape dynamics for patient CH58 involved escape in two regions of the genome.   The two escapes occurred simultaneously and, according to our inference results, affected each other.   To infer such dynamics, high dimensional models that include multiple variants must be considered.   
	
	The results presented for patient CH40 highlight the utility of null models in analyzing HIV escape.  We know that HIV escape occurs through multiple mutation pathways.   Different mutations will likely have different fitness costs and differing levels of escape from MHC presentation and CTL killing.   Multiple variants requiring multiple parameters leads to significant problems related to inference and over-fitting.  Null models can serve a purpose in this context by shifting the question from identification of numerous parameters to identification of variants that are in some way unique and should potentially be the object of further analysis.

	While we have presented a basic framework for computational inference of HIV selective sweeps, there are features of our approach that require more work.  For example, we have not considered the issue of parameter identifiability, see \cite{Miao_2011_SIAM_Review} for a review.  As the number of parameters grows, understanding which combination of parameters can be inferred through a posterior becomes essential.   The computations we have presented occurred on state spaces of modest dimension.  Limiting our analysis to the initial CTL response allowed for this, but future work should extend to the first wave of CTL responses that target $1-3$ epitopes.  While our methods should apply to this greater context, understanding how to construct escape graphs and birth-death processes that can extract biologically useful information in this more computationally challenging setting requires further work.

\clearpage


\newcommand{\noopsort}[1]{} \newcommand{\printfirst}[2]{#1}
  \newcommand{\singleletter}[1]{#1} \newcommand{\switchargs}[2]{#2#1}

\end{document}